\documentclass[fleqn,usenatbib]{rasti}

\usepackage{newtxtext,newtxmath}

\usepackage[T1]{fontenc}

\DeclareRobustCommand{\VAN}[3]{#2}
\let\VANthebibliography\thebibliography
\def\thebibliography{\DeclareRobustCommand{\VAN}[3]{##3}\VANthebibliography}

\usepackage{array}
\usepackage{graphicx}	
\usepackage{amsmath}	
\usepackage{booktabs}
\usepackage{makecell}
\usepackage[capitalise]{cleveref}

\newcommand{\sep}{{\sc sep}}
\newcommand{\pympc}{{\sc pympc}}
\newcommand{\pyephem}{{\sc PyEphem}}
\newcommand{\spalipy}{{\sc spalipy}}
\newcommand{\kadmilos}{{\sc kadmilos}}
\newcommand{\rawtransfer}{{\sc rawtransfer}}
\newcommand{\postgres}{PostgreSQL}
\newcommand{\qtc}{{\sc Q3C}}
\newcommand{\pyehpem}{{\sc PyEphem}}
\newcommand{\astroscrappy}{{\sc Astro-SCRAPPY}}
\newcommand{\django}{{\sc Django}}

\newcommand{\panst}{Pan-STARRS}
\newcommand{\rawinsert}{\mbox{\textit{raw\_insert}}}
\newcommand{\singlesetcreation}{\mbox{\textit{single\_and\_set\_creation}}}
\newcommand{\supercaliblp}{\mbox{\textit{super\_calibration\_la\_palma}}}
\newcommand{\supercalibsso}{\mbox{\textit{super\_calibration\_siding\_spring}}}
\newcommand{\chisqred}{$\chi^2_\textrm{red}$}
\newcommand{\gloss}[1]{\texttt{{#1}}}
\newcommand{\healpix}{\textsc{HEALPix}}
\newcommand{\gotopaper}{Steeghs et al. (in prep)}
\newcommand{\gtecs}{{\sc G-TeCS}}

\title[The GOTO pipeline for transients]{The Gravitational-wave Optical Transient Observer (GOTO) data pipeline and workflow for transient discovery}

\author[J. Lyman et al.]{J.~D.~Lyman$^{1}$\thanks{E-mail: j.d.lyman@warwick.ac.uk (JDL)},
D.~O'Neill$^{2,1}$,
T.~Killestein$^{1}$,
D. Jarvis$^{3}$,
A.~Kumar$^{4, 1}$,
K.~Ulaczyk$^{1}$,
K.~Ackley$^{1}$,
P.~Chote$^{1}$,
\newauthor
M.~J.~Dyer$^{3, 5}$,
M.~Pursiainen$^{1}$,
D.~Steeghs$^{1}$,
B.~Godson$^{1}$,
M.~Magee$^{1}$,
J.~R.~Mullaney$^{3}$,
B.~Warwick$^{1}$,
\newauthor
S.~Belkin$^{6}$,
D.~K.~Galloway$^{6,7}$,
G.~Ramsay$^{8}$,
V.~S.~Dhillon$^{3,9}$,
P.~O’Brien$^{10}$,
K.~Noysena$^{11}$,
R.~Kotak$^{12}$,
\newauthor
R.~P.~Breton$^{13}$,
L.~K.~Nuttall$^{14}$,
B.~Gompertz$^{2}$,
D.~Pollacco$^{1}$,
J.~Casares$^{9, 15}$,
D.~L.~Coppejans$^{1}$,
\newauthor
R.~A.~J.~Eyles-Ferris$^{10}$,
O.~Graur$^{14,16}$,
L.~Kelsey$^{17}$,
M.~R.~Kennedy$^{18}$,
A.~Levan$^{19, 1}$,
S.~Littlefair$^{3}$,
\newauthor
S. Mandhai$^{13}$,
D.~Mata S\'anchez$^{9,15}$,
S.~Mattila$^{12,20}$,
J.~McCormac$^{1}$,
S.~Moran$^{10}$,
C.~Phillips$^{1}$,
K.~Pu$^{6}$,
\newauthor
A.~Sahu$^{1}$,
M.~Shrestha$^{6}$,
E.~Stanway$^{1}$,
R.~L.~C.~Starling$^{10}$,
L. Vincetti$^{8,21,22}$,
E.~Wickens$^{14}$,
K.~Wiersema$^{23}$
\\
$^{1}$ Department of Physics, University of Warwick, Coventry, CV4 7AL, UK\\
$^{2}$ School of Physics and Astronomy, University of Birmingham, Birmingham B15 2TT, UK\\
$^{3}$ Astrophysics Research Cluster, School of Mathematical and Physical Sciences, University of Sheffield, Sheffield S3 7RH, UK\\
$^{4}$ Centre for Electronic Imaging, School of Physical Sciences, The Open University, Walton Hall, Milton Keynes MK7 6AA, UK\\
$^{5}$ Research Software Engineering, University of Sheffield, Sheffield, S1 4DP, UK\\
$^{6}$ School of Physics \& Astronomy, Monash University, Clayton VIC 3800, Australia\\
$^{7}$ Institute for Globally Distributed Open Research and Education (IGDORE)\\
$^{8}$ Armagh Observatory \& Planetarium, College Hill, Armagh, BT61 9DG\\
$^{9}$ Instituto de Astrof\'isica de Canarias, E-38205 La Laguna, Tenerife, Spain\\
$^{10}$ School of Physics \& Astronomy, University of Leicester, University Road, Leicester, LE1 7RH\\
$^{11}$ National Astronomical Research Institute of Thailand (Public Organization), Chiang Mai 50180, Thailand\\
$^{12}$ Department of Physics \& Astronomy, University of Turku, Vesilinnantie 5, Turku, FI-20014, Finland\\
$^{13}$ Jodrell Bank Centre for Astrophysics, Department of Physics and Astronomy, The University of Manchester, Manchester M13 9PL, UK\\
$^{14}$ Institute of Cosmology and Gravitation, University of Portsmouth, Portsmouth, PO1 3FX, UK \\
$^{15}$ Departamento de Astrof\'isica, Univ. de La Laguna, E-38206 La Laguna, Tenerife, Spain\\
$^{16}$ Department of Astrophysics, American Museum of Natural History, Central Park West and 79th Street, New York NY 10024-5192, USA\\
$^{17}$ Institute of Astronomy and Kavli Institute for Cosmology, University of Cambridge, Madingley Road, Cambridge CB3 0HA, UK\\
$^{18}$ School of Physics, University College Cork, Cork, T12 K8AF, Ireland\\
$^{19}$ Radboud University, Postbus 9010, 6500 GL, Nijmegen\\
$^{20}$ School of Sciences, European University Cyprus, Diogenes Street, Engomi, 1516 Nicosia, Cyprus\\
$^{21}$ School of Physics, Trinity College Dublin, College Green, Dublin 2, Ireland\\
$^{22}$ Astronomy \& Astrophysics Section, Dublin Institute for Advanced Studies, DIAS Dunsink Observatory, Dublin, D15 XR2R, Ireland\\
$^{23}$ Centre for Astrophysics Research, University of Hertfordshire, Hatfield, AL10 9AB, UK\\
}
\date{Accepted XXX. Received YYY; in original form ZZZ}

\pubyear{\the\year{}}

\begin{document}
\label{firstpage}
\pagerange{\pageref{firstpage}--\pageref{lastpage}}
\maketitle

\begin{abstract}
Wide-field and high-cadence sky surveys are the first step in the chain of discovery and characterisation of astrophysical transients such as supernovae, kilonovae, and tidal disruption events, each linked to the varied demise of stellar systems. The Gravitational-wave Optical Transient Observer (GOTO) is a telescope array of thirty-two 40\,cm unit telescopes split over two almost antipodal sites. It performs a regular time-domain sky-survey in the optical to $\sim20$\,mag in addition to immediate scheduling of follow-up observations at the locations of external multi-wavelength and -messenger triggers. To facilitate the timely recovery of optical counterparts to these triggers, as well as the presence of serendipitous discoveries of astrophysical transients in the regular sky-survey, a low-latency data pipeline and workflow was developed. The implementation of this workflow is described herein and the quality of GOTO data delivered by it assessed, alongside its performance for prompt transient recovery. Utilising difference image analysis to identify candidate discoveries, the process is typically complete $\sim7$\,minutes after shutter close on the telescope. We further describe later processing of these candidates  -- both automated and human-in-the-loop -- including reporting to the wider community and the triggering of more detailed observations, with a focus on immediate, intra-night characterisation. The workflow is meeting the needs of GOTO to promptly discover, report and characterise infant transients. Nevertheless, areas for further development and improvements are also highlighted. 
\end{abstract}




\section{Introduction}

Very regular patrol of the visible night-sky ($\lesssim 3$\,d cadence) is now possible from a given site to significant depths ($m_\textrm{opt}\sim18-22$\,mag) and with atmosphere-limited spatial resolution. This is largely thanks to major breakthroughs in the manufacture and affordability of large format Charge-Coupled Device (CCD) and  Complementary Metal-Oxide-Semiconductor (CMOS) sensors, as well as the affordability of high-quality optical tube assemblies (OTA) being available at scale. Surveys built around this model are a rich resource for time-domain astrophysics. 

The need for wide-field and high-cadence monitoring of the sky has been amplified following the emergence of immediate, autonomous alerting in the field of multi-wavelength and -messenger time-domain astrophysics \citep[exemplified by][]{gw170817}. This can necessitate rapid coverage of large localisation areas -- delivered in low-latency by high-energy or gravitational-wave observatories -- to pinpoint optical counterparts. Beyond these targeted searches for counterparts, within the broad scope of time-domain astrophysics an ever-widening diversity of optical transients have come to the fore. This outcome is due in no small part to the advent of recent and on-going large-scale synoptic time-domain facilities such as the All-Sky Automated Survey for SuperNovae \citep{asassn}, the Asteroid Terrestrial-impact Last Alert System \citep[ATLAS;][]{atlas}, BlackGEM \citep{blackgem}, the Large Array Survey Telescope \citep{last}, the Panoramic Survey Telescope and Rapid Response System \citep{pansstarrs}, the Wide Field Survey Telescope \citep{wfst} and the Zwicky Transient Facility \citep[ZTF][]{ztf}. The newly uncovered behaviour of these transients, along with unprecedented populations of more common-place variants, is continually challenging our understanding of the demise of stars: the explosive endpoints of stellar evolution. The prospects for further discovery are encouraging with upcoming facilities including the Vera C. Rubin Observatory Legacy Survey of Space and Time \citep[LSST][]{lsst}, the La Silla Schmidt Southern Survey and the Nancy Grace Roman Space Telescope \citep{roman} set to probe new regions of luminosity, time-scale and cosmological distance for transients, coupled with overwhelming numbers of discoveries.

The Gravitational-wave Optical Transient Observer (GOTO) is a recently-commissioned two-node telescope array, building on a prototype design \citep{goto_prototype}, with a principal science focus of obtaining rapid optical electromagnetic constraints for a variety of multi-wavelength and multi-messenger events. The wide-field of view and robotic operations of the constituent GOTO nodes enables quick and automated reaction to external triggers, even with coarse localisation. Between observation campaigns triggered by these external triggers, a routine all-sky survey prioritising more even coverage of the whole sky is performed. GOTO acts complementarily to the ongoing surveys listed previously and adds crucial longitudinal coverage (\cref{sec:goto}) to the global effort for discovering electromagnetic transients in their infancy, during which follow-up observations can be at their most diagnostic for understanding the nature of their explosions and progenitors. Furthermore, as GOTO's primary science aim is reacting to multi-messenger and multi-wavelength, they occupy a consequently dominant portion of the observing mode and strategy. This is in some contrast to other wide-field facilities, which typically require more tensioning of their observing priorities with other science goals surrounding time-domain astrophysics. GOTO is therefore able to react to alerts not routinely covered by other surveys promptly, through a combination of relaxed triggering criteria to follow more marginal GW alerts (\citealt{goto_o3a}; Ackley, Dyer et al. in prep), and coverage of relatively high-rate and large localisation triggers such as \textit{Fermi} Gamma-ray Burst Monitor alerts \citep{fermi_gbm} with location uncertainties of tens of degrees, as seen in \citet{belkin2024} and \citet{kumar2025}.

With wide-field sky-surveys acting as transient discovery machines, an emphasis on rapid dissemination of new discoveries to the community has permitted expedited characterisation of these discoveries during their crucial infant stages.
This systematic excavation of the short-timescale parameter space has revealed previously unseen behaviour -- both intrinsically rapidly-evolving transients, and short-lived phenomena in the infant stages of more typical supernova. 

This paper describes the immediate data processing and transient discovery workflow operations for GOTO facility: from shutter close of the telescope to announcement to the community and further characterisation by other facilities.

\section{GOTO Overview}
\label{sec:goto}

The GOTO project design, hardware, general performance characteristics, and observing strategies are fully described in \gotopaper\ \citep[see also][]{dyer2024}. Here we briefly overview the facility, and in \cref{sec:goto_observing_strategy} detail the observing strategy and survey modes conducive to transient discovery.

GOTO collectively describes a telescope array composed of a large number of small-aperture unit-telescopes (UTs) co-located on a small number of large mounts. There are two nodes -- one in La Palma, Canary Islands ($28^\circ 45^\prime 36^{\prime\prime}\,\mathrm{N},
17^\circ 52^\prime 45^{\prime\prime}\,\mathrm{W}$) and one in Siding Spring, Australia ($31^\circ 16^\prime 24^{\prime\prime}\,\mathrm{S},
149^\circ 03^\prime 51^{\prime\prime}\,\mathrm{E}$). Each node comprises two mounts, with eight UTs on each mount, for a total of 32 UTs. The individual field of view (FoV) of each UT is a little over 5\,deg$^2$, and they are arranged into a $4\times2$ rectangular grid of minimally-overlapping FoVs per mount, giving a mount FoV of $\sim45$\,deg$^2$. UTs are equipped with $\sim$50\,megapixel ($\sim8000\times6000$\,pixels) FLI ML50100 CCD sensors with $1.3^{\prime\prime}$/pixel plate-scale. The UTs each have a filter wheel containing a Baader LRGB filter set.

Each of GOTO's mounts (and the parent domes) are operated autonomously and independently by separate instances of the control software, \gtecs\ \citep[GOTO Telescope Control System; ][]{gtecs}, whilst a central scheduler dictates the optimal next observation(s) for the facility as a whole.
Data from each UT are first written to files on local disks at each node, whereupon they are immediately transferred to a central data centre in the UK for processing and long-term storage.

A table of specific relevant terms and their meaning in the context of discussing GOTO operations and pipeline in subsequent sections is provided in \cref{tab:nomenclature} for reference. Terms with long-established meaning (e.g. image types like \gloss{bias}, \gloss{science}) are omitted.

\begin{table}
 \caption{Definitions of terms used to refer to data types and products.}
 \label{tab:nomenclature}
 \begin{tabular}{lp{0.75\linewidth}}
  \hline
   Term & Description \\
   \hline
   \gloss{tile} & A pointing location within a fixed grid covering all-sky, which is defined by the footprint of an individual GOTO array (\cref{sec:goto_observing_strategy}).\\
   \gloss{event} & An external trigger causing specific \gloss{tile} observations to be made (\cref{sec:goto_observing_strategy}).\\
   \gloss{single} & A reduced image from an individual exposure (\cref{sec:single_image_generation}).\\
   \gloss{set} & A stack of \gloss{single} made from multiple exposures taken at a single pointing (\cref{sec:set_image_generation}). \\
   \gloss{template} & A historical \gloss{set} image taken with a larger number of exposures, and identified as being of good image quality.\\
   \gloss{difference} & A \gloss{set} image after subtraction of a corresponding \gloss{template} (\cref{sec:difference_image_analysis}).\\
   \gloss{super} & A large stack, typically monthly, of \gloss{bias}, \gloss{dark}, or\gloss{flat} calibration frames used for the low-latency processing (\cref{sec:supercalibcreation}).\\
   \gloss{columntrap} & A specific type of \gloss{super} calibration image made of a large number of \gloss{single} images with low sky-background that identifies steps in background level along CCD columns (\cref{sec:columntraps}).\\

  \hline
 \end{tabular}
\end{table}

\subsection{Observing strategy}
\label{sec:goto_observing_strategy}

GOTO generally observes within a fixed grid on the sky, with the grid size dictated by the footprint of a single GOTO mount. These pointings are referred to as \gloss{tiles} and the celestial sphere requires 1024 \gloss{tiles} to be fully covered \citep{dyer_thesis}.\footnote{\url{https://github.com/GOTO-OBS/goto-tile}} The use of a fixed grid of positions simplifies the difference image analysis steps and additionally allows us to treat the individual mounts of GOTO as a single telescope since the provenance of an exposure is only relative for the data reduction steps. The telescope scheduler operates on this assumption, such that a given \gloss{tile} is valid for all mounts and, assuming it is the next highest ranking tile to be observed, is picked-up on a first-come-first-served basis by the mounts at a given site.

Observations of \gloss{tile} pointings during regular operations are made in two modes, namely the fall-back `Sky-survey', and `Follow-up'. The former is GOTO's default observation strategy when no higher ranking observation types are available. Specific \gloss{tile} pointings are ranked higher based on a combination of when they were last observed, and their current airmass, but otherwise provide an unbiased survey covering the whole sky \citep[for full details, see][]{dyer_scheduling}. Within the `Sky-survey', groups of $4\times45$\,second exposures are taken per pointing and have a unique running integer value (\texttt{setnum}) assigned to them. These exposures are individually reduced and analysed into \gloss{single} images (\cref{sec:single_image_generation}), before being combined into a \gloss{set} image for increased depth (\cref{sec:set_image_generation}). 

GOTO continually listens to VOEvent streams of various multi-messenger and -wavelength alerting facilities \citep{gtecs}\footnote{\url{https://github.com/GOTO-OBS/gtecs-alert}} -- \gloss{events} in the GOTO scheduler parlance -- which are formulated into observing campaigns using pre-built strategies depending on the origin and exact nature of the alert. These `Follow-up' observations will trump the `Sky-survey' and insert a series of higher ranking \gloss{tile} pointings, typically with a higher cadence. Indeed different alerts can also override other follow-up, depending on the science ranking of the alert. `Follow-up' usually comprises $4\times90$\,second exposures per pointing, which are reduced and analysed analogously to the `Sky-survey' pointings.

Additionally, the GOTO project began taking \gloss{template} images for each \gloss{tile} from early 2023 when the full hardware implementation was complete. Over $>99$\,per-cent of \gloss{template} images currently used had been assigned as such by mid 2024. These \gloss{template} images are composed of $12\times90$\,second \gloss{set} exposures, significantly deeper than the usual all-sky survey $4\times45$\,second or GRB/GW follow-up $4\times90$\,second stacks, but otherwise processed identically. The use of the \gloss{template} images in difference imaging is detailed in \cref{sec:difference_image_analysis}.

Observations taken as part of transient `Sky-survey' and `Follow-up' are taken in the wider $L$ filter ($\sim400-700$\,nm). This filter is currently used exclusively during transient-related observations as it offers the benefit of higher throughput, allowing GOTO to achieve more sensitive image-depths, at the cost of colour information, which is typically delegated to external facilities to acquire. This choice also simplified \gloss{template} image build-up (since they only need to be acquired in $L$), allowing GOTO to be online and effective for transient discovery faster.

The longitudinal distribution of the GOTO-nodes means that there is no time when both sites are making night-sky observations. However, their latitudinal positions means a region of sky with declination $-20 < \delta < 20$ is accessible at relatively low-airmass from both sites. This region can consequently benefit from added cadence and indeed real-time follow-up campaigns with external facilities aimed at identifying infant transients have exploited this, such as the GOTO Fast Analysis and Spectroscopy of Transients (GOTO-FAST) survey \citep[][Godson et al. in prep]{gotofast_astronote}.

\section{Hardware Implementation}
\label{sec:hardware_implementation}

All processing is performed on a GOTO-dedicated data centre at the University of Warwick, UK, made from a mix of computing and high-capacity storage servers. Some of the computing servers are dedicated to other aspects of GOTO operations. Each computing server used by the pipeline has 32 physical CPU cores (virtually split to provide 64 independent logical processing) and 256\,GB of memory, which allows multiple processes to concurrently perform the typically memory-intensive and sometimes computationally intensive tasks associated with reduction and analysis of GOTO images. Following the telescope control software writing exposures to disk at each site, the data are immediately transferred to this Warwick server for processing (\cref{sec:rawtransfer}). The decision to transfer data immediately instead of processing at the observatory sites was made to ameliorate logistical challenges with performing maintenance tasks on the servers, since no dedicated local staff effort is present on site. Cloud-based hosting was not considered due to concerns over cost, potential volatility of data in the absence of secure long-term funding, and the desire to benefit from working with local scientific computing expertise at Warwick.

A series of six computing servers act as \textit{workers} for the tasks of the pipeline, with two of these also sharing duties of scheduling the tasks and one providing the webserver for pipeline monitoring (see \cref{sec:airflow}). The marshall web framework (\cref{sec:marshall}), and its own task worker, run on a separate, dedicated server. 

Two servers host the database cluster, which serve databases for the pipeline and marshall, alongside other ancillary databases (e.g. contextual information). One server is the primary, with the other a standby which is populated via \postgres\ streaming replication. The standby is a constantly up-to-date read-only copy of the primary database and is the entry-point for user and marshall queries that need only to fetch data, thus lightening the load on the primary, to which the pipeline intensively reads and writes records. The standby is also capable of being immediately promoted to primary, should the primary server undergo failure.

Data are stored locally to the cluster on a rolling retention policy, typically six months' worth of processed image data. This time-window is more than sufficient to ensure required access for further processing, forced photometry etc. is expedient due to high-throughput network links between servers in the server. Raw and processed data are additionally archived shortly after processing to a distinct PB-scale storage server managed by the Scientific Computing Research Technology Platform of the University of Warwick.

\section{Pipeline implementation}
\label{sec:pipeline_implementation}

The pipeline is intended for rapid transient identification and dissemination. Precise photometric calibration is not emphasised and deferred to future data products from GOTO (Jarvis et al. in prep). Following the tradition of overly-grandiose naming for software packages, we dubbed the data reduction and analysis pipeline \kadmilos\ during its development, and will refer to it as such hereafter.

\begin{figure*}
    \centering
    \includegraphics[width=\linewidth]{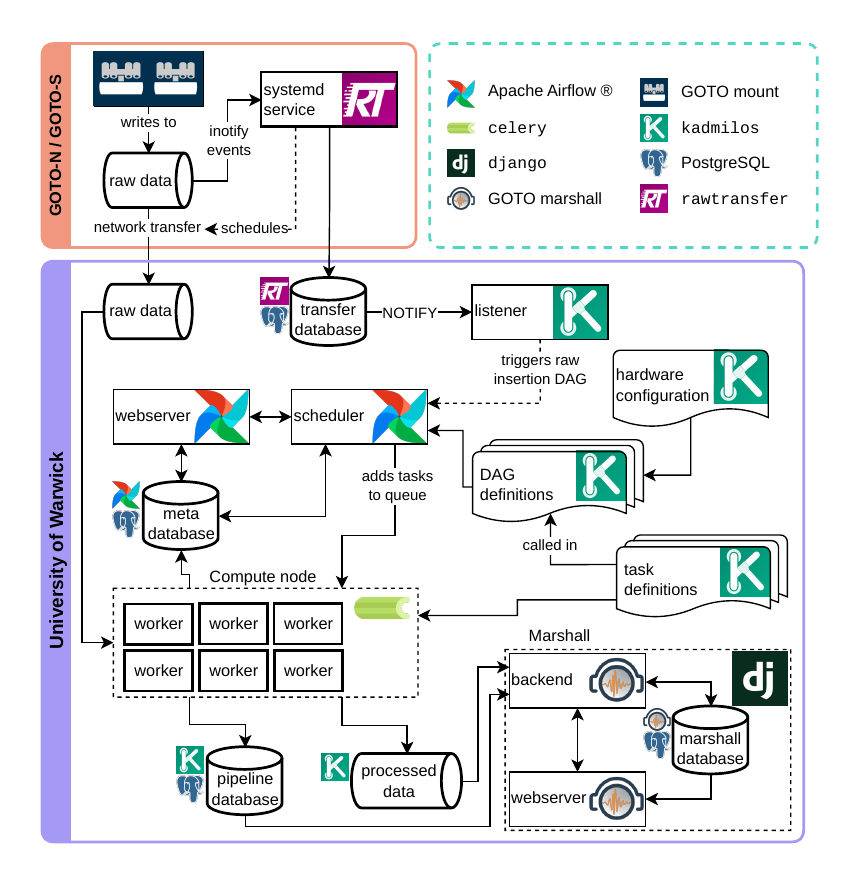}
    \caption{A schematic showing the data flow and software implementation of the GOTO transient discovery workflow. Full details are given in the respective sections. Succinctly, data acquired and written by the GOTO mounts to a local file system at the respective observatory \citep[see][]{dyer_thesis}. A \rawtransfer\ service (\cref{sec:rawtransfer}) at each site monitors these local file systems, packages the new FITS files as they are created and transfers them to a large capacity file system at the University of Warwick, writing the details of the transfer to a record in the transfer database, also held in Warwick. The insertion of the record emits a NOTIFY signal which tells the Airflow (\cref{sec:airflow}) scheduler that a new raw insertion DAG (\cref{sec:dagdetails} should be scheduled to process the file. The tasks of the \kadmilos\ pipeline (e.g. \cref{sec:single_image_generation}) are performed by a cluster of Celery workers, scheduled by Airflow to follow the respective DAG's logic. As part of these tasks, the workers write processed data and results to a central filesystem and database. These central data resources are pulled by the GOTO marshall (\cref{sec:marshall}), where additional metadata such as contextual information is fetched, and source associations are made. The result source data is then visualised, vetted, and reported by collaboration members via a webserver.}
    \label{fig:kadmilos_dataflow}
\end{figure*}

The data processing of astronomical data from a facility like GOTO largely falls under the Extract-Transform-Load (ETL) model prevalent in many other data handling systems \citep{etl}. Succinctly, data from one or more sources is input, undergoes transformation (in this case requiring significant amounts of data reduction and analysis), before storing distilled outputs of the data in such a way that they can be disseminated to users or further processing tools. Practically, components of the pipeline are responsible for taking raw image data from each GOTO UT, and outputting final image reductions, photometric measurements, and new difference image sources. We here detail various aspects of the implementation of \kadmilos, including the associated solutions for transferring data, monitoring the processing, and storing the output. An overview schematic of the data and software flow is shown in \cref{fig:kadmilos_dataflow}.

\subsection{Transferring raw data from the observatories}
\label{sec:rawtransfer}

As the computing infrastructure is held at the University of Warwick, raw data in the form of FITS files\footnote{\url{https://fits.gsfc.nasa.gov/fits_standard.html}} must be copied immediately from each site upon writing before being picked up by \kadmilos. A small package \rawtransfer\footnote{\url{https://github.com/GOTO-OBS/rawtransfer}}, builds around \texttt{ssh} data transfer, for this purpose. Additional safeguards are in place to ensure the uniqueness of file transfers, error handling for invalid FITS files, retries of transfers in the case of, e.g., temporary network down-time, and a database record of each successful transfer. A copy of \rawtransfer\ runs locally at each site and a monitoring thread listens for filesystem events from the Linux kernel \emph{inotify} subsystem indicating new a raw data file was written by the telescope software \gtecs. The file is then queue-scheduled for transfer to Warwick, with multiple worker threads processing the queue in sequence. In the absence of network issues, this queue processing is effectively immediate.

Using a \postgres\ database for the ground-truth of transfers allows for the creation of triggers (functions executed when a specific operation occurs, e.g. insertion of a new row into a table) which themselves emit a NOTIFY message. External services can listen to these NOTIFY messages. \kadmilos\ does exactly this for the purpose of initiating the pipeline processing cascade for the transferred file. The 50th and 95th percentile timings for FITS checking and transfer from the La Palma site to Warwick are $P_{50} = 9$ and $P_{95} = 31$\,seconds post-file-writing. The La Palma node benefits from a dedicated level 2 VLAN connection to Warwick. For the Siding Spring site, we are at the mercy of a dynamic route over a public network between gateway machines at Warwick and Australia National University, reflected in slower transfer timings: $P_{50} = 21$ and $P_{95} = 207$\,seconds. The typical size of a \gloss{raw} science frame transferred is comparatively small at 42\,MB since the integer data extension can be stored efficiently.


\subsection{Apache Airflow: Orchestrating \kadmilos}
\label{sec:airflow}

A primary concern with deciding on the pipeline implementation was a robustness of tasks execution on different schedules, the ability for multiple work-flow patterns to be decided on-the-fly, the means to easily re-process (or partially re-process) batches of data, and a monitoring system to identify issues as they may arise. 

Apache Airflow\footnote{\url{https://airflow.apache.org/}} (Airflow) well suited these requirements, with the added bonuses of being an open source project implemented in Python, the language also used for \kadmilos. Airflow is particularly adept at defining ETL pipelines, and offers easy horizontal scalability that is attractive for a project like GOTO where we require immediate processing of significant chunks of data from 16 UTs at $\sim$minute intervals almost continually during good weather at both sites.

Workflows in Airflow are designed as directed acyclic graphs (DAGs) of tasks, where each task can have a conditional dependency on one or more upstream tasks. A specific processing instance of a workflow is referred to as a DAG run, which comprises one or more task instances.
Airflow comprises four main components, a metadata database, one or more schedulers, one or more workers, and a web server. Extensive documentation is available on the Airflow webpages, and so we only briefly summarise the components here.

The metadata database is a relational database created by Airflow to allow it to monitor the processing environment and its own actions, as well as keep a record of all historical actions such as initiating DAG runs, and the outcome of task instances. We implement it using \postgres\ for our deployment, which allows for multiple schedulers to be run concurrently.

The scheduler makes primary use of the metadata database, inserting and updating records to journal and monitor the status of all task instance and DAG runs -- deciding what needs to be executed next on a regular interval. For example, it identifies which tasks are currently due for processing and passes those to the workers via a task queuing system, marks downstream tasks of any recently completed tasks as requiring to be processed, and updates the status of DAG runs based on outcomes of their child tasks. Additionally the scheduler checks the outputs of the workers to update whether tasks are currently being processed, have successfully completed, have failed etc. A scheduler instance naturally runs on a cycle of operations, each taking finite time. To improve the prompt scheduling of tasks and increase their throughput to be picked up by the workers, we run two scheduler instances on separate compute servers. The use of \postgres\ as the metadata database allows for record-level locking and removes collision and deadlock potential from having multiple schedulers access it pseudo-concurrently. The scheduler is also responsible for parsing the files that define the workflows (or DAGs), such that their order of task dependency and scheduling requirements can be stored in the metadata database.

The workers are the processes that process tasks defined in a workflow (or DAG). Storing \kadmilos\ on a shared filesystem in the data-centre allows each worker server to see identical code for the parsing and execution of DAGs and their constituent tasks, avoiding any conflicts that may arise during updates to either. Airflow, by default, makes use of Celery\footnote{\url{https://docs.celeryq.dev/en/stable/}} as an executor of tasks. Celery is a distributed task queue, which takes care of ensuring the worker processes pick up, execute, and report back on tasks. Many worker processes can be spawned on a given server, with distribution to multiple servers also readily supported. The workers update the metadata database once tasks are completed (and if they failed, succeeded etc.), which the scheduler will query on its next iteration to determine the next course of action for each DAG.

Lastly, the Airflow web server provides an informative UI for interrogating the status of the scheduler, workers, and DAG/task processing. It also provides a user interface to the Airflow API, allowing the user to manipulate the processing of DAG runs and tasks, especially useful for re-running previously failed tasks due to a transient issue (e.g. network or filesystem). The web server is routinely checked by operations teams to assess the health of the system, and DAG run failures are propagated to the GOTO internal communication platform (Slack) for assessment.

\subsubsection{DAG details}
\label{sec:dagdetails}

\kadmilos\ uses both triggered DAGs and scheduled DAGs. The latter run at specific intervals or at a given time of day. For the former either an external process, or another DAG, is responsible for initiating a DAG run. 

The primary triggered DAGs are named \rawinsert\ and \singlesetcreation. Upon insertion of a record in the \rawtransfer\ database (\cref{sec:rawtransfer}), due to the successful transfer of a FITS file, a notification is issued with a payload including the file path of the newly transferred file. This payload is picked up by a dedicated listener (running as a systemd service) which triggers the \rawinsert\ DAG using the payload. The cascade of \kadmilos\ processing for that file then ensues. For \gloss{science} frames the \singlesetcreation\ DAG is then triggered after some rudimentary quality checking of the data. This latter DAG takes the raw data through the steps detailed in \cref{sec:basicccdprocessing,sec:single_image_generation,sec:set_image_generation,sec:DIA}.

The primary DAG running on a schedule is the creation of \gloss{super} calibration frames at each site: \supercaliblp, \supercalibsso{} (\cref{sec:supercalibcreation}).

The above are the DAGs most pertinent to the transient workflow in GOTO, although additional DAGs for housekeeping as well as deprecated workflows (e.g. forced photometry) are visible in the public repository.\footnote{\url{https://github.com/GOTO-OBS/kadmilos-public/tree/main/kadmilos/dags}}

DAGs are written to be intentionally idempotent. This naturally facilitates the ability to re-process (or `back-fill') data. (Non-astrophysical) transient issues such as an inaccessible database connection, or compromised data integrity due to a hardware issue, cause DAGs to fail. The web UI or command-line Interface to Airflow allows the GOTO operations team to promptly identify and rerun the failed DAGs once the issue is corrected. Due to the splitting of DAGs into individual component tasks, it is also typically the case that only a sub-set of the processing for each failed DAG needs to be attempted again.

\subsubsection{DAG versioning}
\label{sec:dagversioning}

Given the slowly changing dimensions of GOTO's configuration (e.g. replacement of cameras) we employ a DAG versioning system based on the notion of a `hardware version' for GOTO. Once a change is made to the system that alters the expected flow of processing and/or invalidates calibration files for a given UT, we generate a new hardware configuration file to describe it, which automatically authors a new version for those DAGs that rely on knowledge of the hardware set-up (i.e. \gloss{super} calibration creation, \cref{sec:supercalibcreation}), and appends the hardware version to the name of the DAG. An example of a hardware configuation file can be found in the frozen public repository of \kadmilos.\footnote{\url{https://github.com/GOTO-OBS/kadmilos-public/blob/main/kadmilos/example.hardware_config_2020_0.yml}} We note that the release of Airflow 3.0 in April 2025 allows for DAG versioning natively in the Airflow ecosystem, and is a planned upgrade to \kadmilos.

As part of this hardware versioning, a camera configuration file is held to store immutable properties of the cameras being used, such as the expected shape of the data, and the pixel scale when mounted on a GOTO UT. This configuration is used to validate raw data and in some processing steps.

\subsubsection{Overall Impressions}

Airflow is not typically employed or designed for very high-throughput, low-latency processing as we are performing with GOTO. Indeed the tool is specifically advertised as such, citing concepts like a non-focus on minimising lags for scheduling tasks (and their subsequent pick-up by workers). Although undoubtedly additional performance could be gleaned from a concerted effort in tweaking of the Airflow parameters and better DAG design (in addition to time savings through optimisation of the tasks themselves), the timescales for processing have met our original aims for the project, and so effort has not yet been directed to this aim. Considering the large and typically continuous wave of DAG run scheduling, in terms of overall throughput, Airflow has performed admirably in the face of our demands. Currently there have been over 13 million DAG runs and 130 million tasks instances fully documented in its metadata database and counting.\footnote{In practice, using Airflow's archiving command, old metadata records are moved to separate tables from the ones frequently queried by the scheduler. We employ this as a best practice but saw no significant degradation of scheduler performance prior to this.} The orchestration by Airflow rarely encounters issues, except those externally enforced, and scheduling and execution of tasks maintain the same performance level as it did initially.

\subsection{\postgres: Persistent data products storage}
\label{sec:postgresql}

Alongside the Airflow metadata database, the choice for data products from \kadmilos\ is also a \postgres\ database, in the same cluster.
The database contains schemas holding: raw, calibration, and image data; photometry; scheduler events and surveys; externally-derived catalogues. 

For photometry tables, \qtc\ spatial indices \citep[Quad Tree Cube;][]{q3c_paper,q3c_ascl} are added on the RA and Declination columns to allow for performant coordinate-based querying (e.g. cone or polygon searches). A copy of the ATLAS-REFCAT2 catalogue \citep{atlasrefcat2}, used for photometric calibration (\cref{sec:single_photometric_calibration}), is also held locally in a table of the database, for which \qtc\ spatial indexes were also generated.
The photometry tables are by far the largest, to the point where partitioning significantly helps with their management and querying. These partitions are done on the \texttt{date\_mid} timestamp value, with each photometry table holding one calendar month of records. Cone searches of $\lesssim1$\,arcmin take $\sim0.1$\,seconds to search 1\,year of single photometry partitions -- $\mathcal{O}(10^{11})$\,records.

\subsection{Basic raw data processing}
\label{sec:basicccdprocessing}

The following subsections describe the CCD reduction applied to every new GOTO \gloss{science} frame in low-latency, after ingestion into \kadmilos.

Other image types of calibration frames (\gloss{bias}, \gloss{dark}, \gloss{flat}) are ignored during immediate low-latency processing as they are not needed individually. Instead, these frames are batch processed into \gloss{super} calibration frames (\cref{sec:supercalibcreation}), for use in the reduction of future \gloss{science} frames. 

\subsubsection{Pre-reduction corrections}

The edges of the CCD arrays -- $\sim10$\,pixels along each edge \citep{dyer_thesis,goto_prototype} -- are typically compromised due to detector edge-effects and increased noise level. These regions are therefore trimmed from all frames. The median value in an overscan region (removed during prior trimming) is then subtracted from all remaining pixel values, to account for the hardware bias level. The data are then multiplied by a gain value ((e$^-$/count) for the specific camera channel, to convert the pixel values from counts to electrons. Gain values for every camera were determined for each of the two readout channels of the sensors, with the mean and variance of a set of differently-illuminated flat-fields used to calculate them.

\subsubsection{Generation of super calibration frames}
 \label{sec:supercalibcreation}
 
Regular CCD processing involves the removal of pixel-to-pixel pattern noise in the bias level, and any dark current contribution to the pixel values. Pixel-to-pixel sensitivity to incident photons and an overall illumination correction are needed to further standardise the pixel values across the CCD. Finally, dead, hot, or otherwise non-linear, pixels must be identified to prevent their erroneous readings propagating into the analysis. The sensors in GOTO are maintained at $253$K during operation, resulting in only a modest dark current. Nevertheless, the fraction of non-linear and hot pixels is elevated above that of high-grade, cryostat enclosed CCDs. Full details of the sensor operation will appear in Steeghs et al. (in prep).

To correct for these effects, \kadmilos\ generates monthly banks of \gloss{super} calibration frames. These frames are heavily-stacked versions of individual calibration frames, produced per camera for those within the current hardware configuration (\cref{sec:dagversioning}). Images of type \gloss{bias}, \gloss{dark}, and \gloss{flat} are taken on a regular schedule each evening and morning by the GOTO control software \citep{gtecs}. A scheduled DAG spawns tasks to search a maximum of 3\,months previous to fetch up to 80 observations of each image type, starting from the most recently acquired. If fewer than 50 of any type is found, the \gloss{super} calibration creation is skipped for that camera (and low-latency processing will fall back to the previous month's calibration images). Individual frames are combined pixel-wise with sigma-clipping of pixels. No weighting is applied to individual frames during combination. For \gloss{dark} frames, the combination is performed per unique camera--exposure-time combination. Exposure time subsets are 30, 45, 60, or 90\,seconds as these match the possible choices used for \gloss{science} images in GOTO's observing strategy. Data are scaled within these subsets by their exact exposure time. The \gloss{flat} frame combination is performed per unique camera--UT--filter combination, and is preceded by scaling each frame by its respective inverse median. 

Some crude quality assessment is performed on individual calibration frames when they are initially read, checked and stored by \kadmilos\ during the \rawinsert\ DAG. For \gloss{bias} and \gloss{dark} (and also \gloss{science}) image types, these checks solely identify obviously problematic data that would fail during processing, e.g. they ensure the background level is within a sensible range and the shape of the data array is as expected for the camera. For \gloss{flat} frames, however, the pixel-value distribution is further analysed as part of these checks. The skew and kurtosis values of this distribution were found to be a strong indicator of both the presence of cloud cover, and the abundance of astrophysical objects in the frames (due to extended exposure times during dark conditions). As each of these effects compromises the quality of the flat, we set empirically-defined limits on these numbers. Any frame lying outside these limits is appropriately flagged during the database ingestion stage, such that it is not considered for inclusion in any \gloss{super} \gloss{flat} frame stack.

The \gloss{super} \gloss{bias} frames are simply subtracted from the \gloss{science} data during reduction, but the \gloss{super} \gloss{dark} and \gloss{flat} frames additionally allow for the identification of misbehaving pixels. This task calls for the additional creation and storage of a mask alongside the parent calibration data. Hot or dead pixels are identified in the \gloss{dark} frame stack by calculating the ninety percentile range ($P_{90}$) of pixel-wise distribution of values in the constituent individual frames. Pixels in the \gloss{super} \gloss{dark} frame that have a $P_{90}$ which is itself more than six standard deviations from the distribution of $P_{90}$ values of all pixels are flagged due to erratic behaviour. Further, the highest 0.1\,percent of pixel values are additionally flagged as the population of hot pixels. Non-linear pixel responses are identified through ordering the individual \gloss{flat} frames by their median pixel value. The first and last quartiles of frames ordered thus are separately combined. A `ratio' frame is calculated from a division of these stacks, which is modelled with a smoothly-varying background using the same procedure as done for \gloss{science} frame reduction (\cref{sec:single_background_removal}). Any pixels in the background-subtracted `ratio' frame that are more than $7\sigma$ deviated based on the local background root mean square (RMS) are flagged.

Currently, \gloss{super} calibration frame creation is a trusted process -- i.e. newly-created files are employed in low-latency image processing immediately after generation. Significant failures are not typically seen owing to the selection of a large number of individual calibration frames from a significant time-period (alongside quality control of \gloss{flat} frames). A more secure method, however, is envisaged whereby manual inspection of the \gloss{super} calibration frames is performed, prior to them being made active by the pipeline. The lack of poor performance by the \gloss{super} stacking routines has de-prioritised adding in this stage so far.

After applying these standard CCD corrections to \gloss{science} frames, a remaining feature was seen in the data from manual inspection during development of the pipeline. The feature manifests due to trapped charge in columns from defective pixels, which calls for an additional \gloss{super} calibration frame to be produced and used in our reduction recipe (see next section).

\subsubsection{Identifying and correcting column traps}
\label{sec:columntraps}

\begin{figure}
    \centering
    \includegraphics[width=\linewidth]{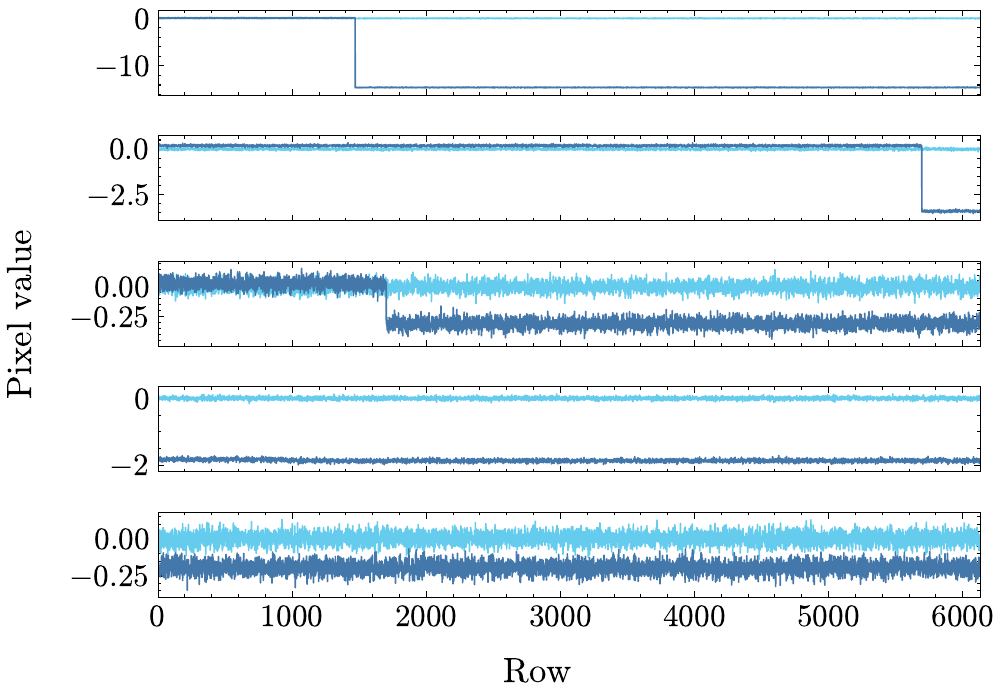}
    \caption{Column-wise traces of pixel value in deep-stacked single images of a GOTO camera highlighting various column-trap behaviour. Each panel shows the trace for a visually identified bad column (dark blue), alongside the trace for the neighbouring, good, column (light blue). The top three panels show `step' features in the traces of varying depths, with the second panel also showing some offset \textit{above} that of the neighbouring good column. The bottom two panels show overall offsets from the good trace along the entire column length.}
    \label{fig:column_trap_individual_bad_columns}
\end{figure}

In the transfer of charge along the readout axis (columns, in this instance) of a CCD, electrons can become `stuck' in the semi-conductor pixel wells. This trapping can cause a partial, or entire loss of signal along the remainder of the column after the misbehaving pixel, commonly referred to as column traps or charge traps. Column-wise traces of pixel values are shown in \cref{fig:column_trap_individual_bad_columns}. To create these traces, a 700\,pixel wide region of 500 randomly chosen \gloss{single} images (i.e. \gloss{science} frames that have been bias-, dark- and flat-corrected) taken during February 2025 by GOTO-2 UT2 were collected. (This sensor, and the specific region highlighted in the figure, were chosen as a region of elevated column trap density to illustrate the method.) These data were then combined, pixel-wise, using a weighted (inverse background variance) mean with sigma-clipping. These frames were selected for having a low sky-background level of $<400$\,counts since bright sky conditions make the identification of these (often subtle) bad column behaviours difficult, and had their backgrounds removed (see \cref{sec:single_background_removal}) prior to combination. Manual inspection of the stacked data was used to identify the columns displayed.

The functional form of the traces seen can be variably best described as a constant, or a constant with a step function acting over 1\,pixel. A series of model fitting is therefore performed per-column using {\sc lmfit} to find the optimal description of the trace and identify a pedestal level which can correct each column of the CCDs, as follows.

Firstly, a constant model is fitted to each column using least-squares. Since only a small fraction of columns display a column trap, these initial constant model fits are used to identify them under the assumption that most columns should be well-described by the constant model. To lessen the effect of surviving hot or dead pixels in the stacked image, the constant model is fitted to a median-filtered version of each row using a kernel size of five pixels. Per image, a conservative constraint indicating a poor fit is established as:

\begin{equation}
\label{eq:chi2thresh}
    \chi^2_\textrm{red, thresh} > \tilde{\chi}^2_{\textrm{red}, i} + 8 \times \textrm{med}(|\chi^2_{\textrm{red}, i} - \tilde{\chi}^2_{\textrm{red}, i}|),
\end{equation}

where $\chi^2_{\textrm{red}, i}$ is the reduced chi-squared of the $i$th column, and an overhead tilde indicates the median of the corresponding set. A separate constraint on the level of the constant model for the $i$th column, $C_i$, is also established:

\begin{equation}
    C_\textrm{thresh} < \tilde{C}_i - 2 \times \textrm{med}(|C_i - \tilde{C}_i|).
\end{equation}

Any constant model satisfying either or both of these criteria is put forward as a candidate column trap. The first and last 4 columns of the CCD are not considered, as edge effects introduced in reduction resulted in these being typically always flagged as candidates.

A one-step model is then fit to each candidate column. The step is forced to act over an infinitesimal length, to mimic the discrete change in level from one pixel to the next. As such a function cannot be differentiated, differential evolution is used for the fitting method \citep{differentialevolution}. As for the constant model, the threshold of \cref{eq:chi2thresh} is applied to the \chisqred values of the one-step models. For any columns still meeting this criterion, a two-step model is fitted, allowing for two discrete jumps in the pixel value trace. Where this two-step model produces a better \chisqred value than the one-step model, the two-step model is used.

\begin{figure*}
    \centering
    \includegraphics[width=\linewidth]{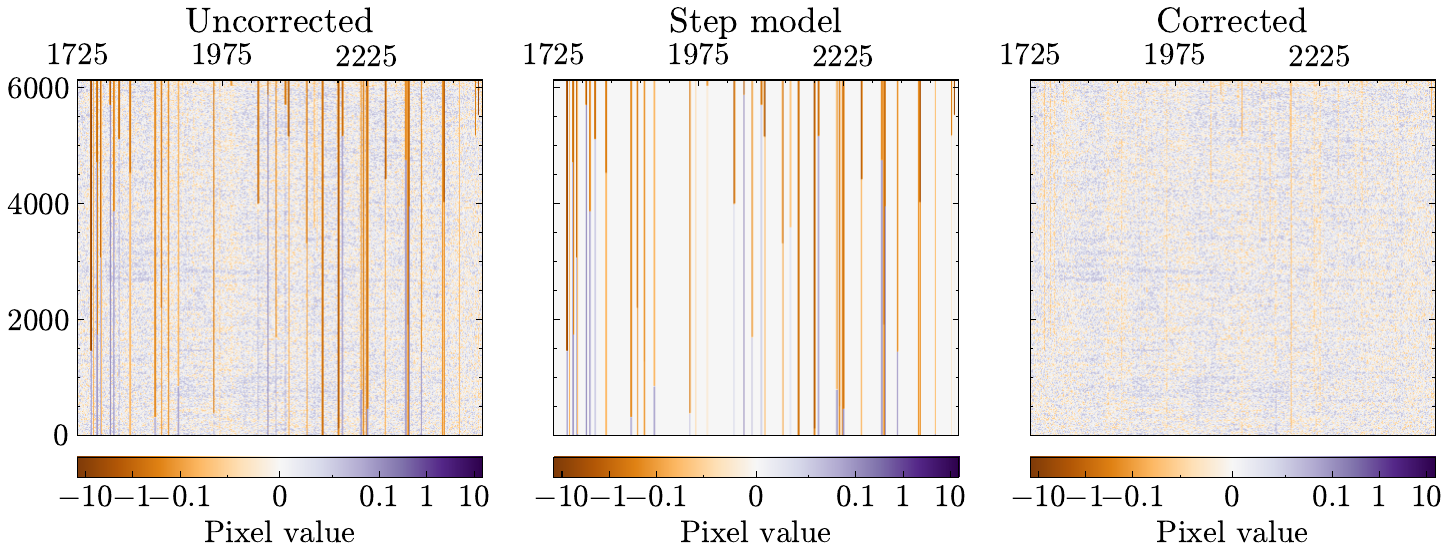}
    \caption{The use of a column-wise step-model to correct column charge trap defects in GOTO data. Left: A deep stack of 500 single science images prior to the correction. Middle: The step-model of the column traps following the procedure described in the text (\cref{sec:columntraps}). Right: Subtraction of the step-model from the uncorrected data showing the removal of almost all trap features. Very subtle and or small-length traps remain, visible due to aggressive visualisation scaling. The read-noise alone in a given GOTO image dominates over the amplitude of these features. Axis values relate to pixel coordinates and the colour scale employs a broken log-linear scheme to enhance contrast at the low background level.}
    \label{fig:column_trap_step_model}
\end{figure*}

A final \gloss{columntrap} calibration frame is produced using these one- and two-step model values, and filled with zero elsewhere. The process is visualised in \cref{fig:column_trap_step_model}. In regular creation of \gloss{super} \gloss{columntrap} calibration frames \kadmilos\ uses between 120--200 single images since it was found stacking much beyond $\sim120$ single science images gave negligible changes in the calibration output. The addition of this step model cleanly removes those columns with strongly-defective charge transfer -- the difference in pixel level from neighbouring well-behaved columns is small -- ($\mathcal{O}(0.1\,\textrm{e}^-)$ -- compared to the read-noise level of a single image -- $\mathcal{O}(10\,\textrm{e}^-)$. Of the marginally-defective columns that are undetected in the step model fitting (right panel \cref{fig:column_trap_step_model}), their effect does not contribute significantly to the uncertainty budget for photometry. These subtly defective columns are apparent in \cref{fig:column_trap_step_model} only owing to the use of a stack of 500 \gloss{single} frames, but are undetectable in any given exposure as the read-noise dominates.

When applying the \gloss{columntrap} correction during data reduction, a final check is made: there can be scenarios where the background level of a \gloss{science} frame is smaller than the step in the calibration frame. To avoid imprinting negative versions of strong traps during reduction of these frames, a limit is placed on the correction value to be at most the level of the local background (calculated in the fashion of \cref{sec:single_background_removal}).

\begin{figure}
    \centering
    \includegraphics[width=\linewidth]{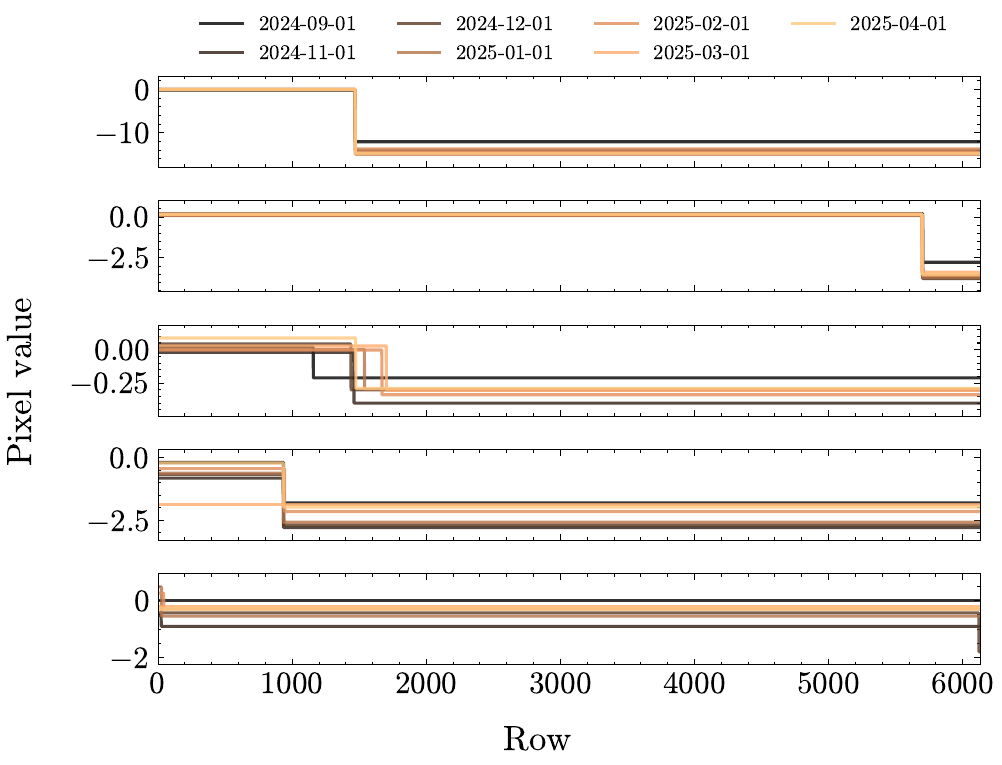}
    \caption{Column-wise traces of pixel value in \gloss{columntrap} calibration frames showing the evolution of column-trap behaviour. Each panel shows the trace for the same columns as shown in \cref{fig:column_trap_individual_bad_columns}, with the colour indicating the date the calibration frame was created.}
    \label{fig:column_trap_evolution}
\end{figure}

We found that the evolution of the traps is non-negligible, albeit slowly evolving (\cref{fig:column_trap_evolution}). Monthly creation of \gloss{columntrap} calibration frames has proven a good compromise to capture this evolution for use during the low-latency image processing, without unnecessarily creating them too frequently.

\subsubsection{Application of super calibration frames}
\label{sec:application_super_calibration}

During low-latency processing of a \gloss{science} frame, the appropriate \gloss{bias}, \gloss{dark}, \gloss{flat} and \gloss{columntrap} frames are found -- i.e. those created most recently to the observation date of the \gloss{science} frame,\footnote{\kadmilos\ further enforces these calibration frames must be historic (pre-observation), in the case of reprocessing a \gloss{science} frame.} and which match camera-UT-filter combination, as required.

The corrections are applied in the typical CCD reduction manner \citep[e.g.][]{ccdreduction}: subtraction of the \gloss{bias}, \gloss{dark}, and \gloss{columntrap} frames, before division of the normalised \gloss{flat}. Pixels identified as hot/dead/non-linear (\cref{sec:supercalibcreation}) are propagated in a bit-wise fashion to generate an initial bad pixel mask for the now-reduced \gloss{science} frame. Columns corrected by the \gloss{columntrap} calibration frame are not identified in the mask since their pixels are sufficiently well-recovered and behaving to be useful in photometric measurements. At this point the reduced \gloss{science} frames are ready for data processing.

\subsection{Creation and science processing of \gloss{single} images}
\label{sec:single_image_generation}

A \gloss{single} image is a processed and analysed \gloss{science} frame produced by a single exposure of a GOTO UT. The creation and processing of CCD-reduced data (\cref{sec:application_super_calibration}) is described in the following sub-sections.

Unless stated, where values are calculated and/or visualised in this and subsequent sections, they are taken from GOTO images taken in the time period 2025-07-28 to 2026-01-28.

\subsubsection{Cosmic-ray removal and bad-pixel correction}
\label{sec:single_cosmic_ray_bad_pixel_correction}

Cosmic-rays originating from the high energy particle background pose as random hits on CCD detectors causing a localised spike in signal. Through the difference image process (\cref{sec:difference_image_analysis}), these can manifest as positive detections, unnecessarily burdening the process of identifying real astrophysical transients. The sharp profile of cosmic-ray detections typically allows for their robust identification in over-sampled ground-based astronomical imaging, with the most popular technique -- convolution of the data with a Laplacian-like kernel \citep{vandokkum2001} -- able to also perform well in critically-sampled data. Given the pixel size of GOTO and the optical performance, under good conditions at the observatories the PSF can enter the regime of critical sampling.
The algorithm used is a direct implementation of the \citet{vandokkum2001} method, accessed via \astroscrappy\ \citep{astroscrappy}. It is applied iteratively, identifying and masking cosmic-ray hits each time until no new pixels are identified.

The cosmic ray pixel mask is combined with the camera's bad pixel mask (comprised of non-linear and hot/dead pixels identified during the production of calibration frames; \cref{sec:application_super_calibration}). Since one cannot recover useful information from these pixels, it is viable to simply track these pixels via a bad pixel mask and remove their influence from any measurements or photometry. However, if they are left untreated they can cause further unwanted consequences. For example, when transforming data spatially (as needed for difference imaging), interpolation of sharp features onto a new pixel grid can produce aliasing effects, spreading their impact, and they can affect segmentation of the image at the source-detection stage. We thus perform a (largely aesthetic) interpolation over the bad pixels using a 2D Gaussian kernel with a small ($\sim1$\,pixel) sigma value, replacing their value by the value of the kernel. The replacement values are much closer representations of the expected good pixel value, diminishing their adverse effect on further processing (cf. examples given above). Their positions are nevertheless traced via the bad pixel mask, still allowing for them to be accounted for during final photometry. The mask itself is an integer array, allowing for bit-wise flagging of pixels to trace their respective provenance. 

\subsubsection{Background removal}
\label{sec:single_background_removal}

Scattered-light and sky-brightness act as noise contributions to the reduced \gloss{science} frame data, as well as elevating the mean background pixel count above zero (the idealised case).

A typical method to remove a potentially spatially-varying background is to identify the local background on a coarse mesh of points and then interpolate between those points using some smoothly evolving functional-form such as a low-order spline. In practice, the determination of the local background at the mesh points, and the mesh-points themselves, are subject to sigma-clipping to remove outliers, which may be caused by real astrophysical objects. The background calculation is performed by the \sep{} package \citep{sep}, itself using the algorithm of \citet{sextractor}.

A two-step process is employed. Firstly, an initial background formed from a very coarse resolution mesh is used to identify pixels $>2.5\sigma$ above the local background. The pixels are included, along with the frame's bad pixels (\cref{sec:single_cosmic_ray_bad_pixel_correction}), in a mask for construction of a background based on a finer resolution mesh background. This final background allows for the subtraction of the spatially varying sky level from the pixel values, as well as calculation of the spatially varying RMS of the frame, for later use in uncertainty calculation during photometry. The final background mesh is constructed with points separated by $\sim64$\,pixels in each direction (exact mesh size determined by the final image size to ensure a roughly integer number of mesh spacings). Although large nebulae and galaxies will have their extended emission captured (and removed) by this background-resolution, we are motivated to reduce issues for image subtraction (\cref{sec:DIA}) and ensure reliable point-source photometry, rather than preserve extended object morphology.

Due to slight differences in the gain and noise properties of the two readout channels that make up each GOTO camera CCD, this process is performed separately for each channel, before being combined to produce a final background object. The presence of bright stars on the dividing line between the two readout channels can produce unrepresentative discrete steps in the background. However, the locations around bright stars are typically compromised due to scattered light from the PSF of the star in any case. Furthermore, the issues due to bright stars on the intersection of the channels occur much less frequently than issues caused by trying to fit a single, smoothly varying spline interpolation of the background to the discrete change in noise and level properties of the image. 

\subsubsection{Source detection}
\label{sec:single_source_detection}

Point-like sources are identified and measured in each image to enable further calibration, required for transient discovery. The detection routine is that of \sep{} \citep{sep}, also used for background removal (\cref{sec:single_background_removal}), and makes use of the bad pixel mask, and the background RMS to create a de-blended segmentation map \citep{sextractor} of sources with at least 7 pixels above 2.5$\sigma$ of the local background. Following this initial detection pass, cuts are made to remove sources with any of the following:

\begin{enumerate}
    \item a not-finite value for one or more shape parameters.
    \item a flag from \sep{} indicating the photometry is compromised.
    \item a full-width half-maximum (FWHM) value $\lesssim1$\,pixel.
    \item a centre too close to the edge of the field of view (within $1\times$ the average FWHM of the image).
    \item an overall $\mathrm{SNR} < 5$.
\end{enumerate}

Half-flux diameter (HFD) measures the diameter of a circle centred on the source that contains half the total flux. For a Gaussian profile, it is related to the full-width at half-maximum (FWHM) by $\mathrm{HFD} \simeq 0.6 \times \mathrm{FWHM}$. HFDs are calculated for each source, again using \sep{}. The distribution of HFD values is sigma-clipped, before taking a median to obtain a representative HFD for the image. Two aperture radii, ``small'' and ``large'', are then defined as $1$ and $3\times\textrm{HFD}$ respectively. Although the smaller aperture is nominally larger than one which will maximise signal-to-noise (SNR) for a Gaussian profile,\footnote{\url{https://web.ipac.caltech.edu/staff/fmasci/home/mystats/GaussApRadius.pdf}} it ameliorates centroid issues, particularly a concern under good conditions given the relatively coarse pixel-scale of GOTO compared to the delivered image quality. The PSF shape is also unable to be held constant over the wide field of view of the fast UTs, guaranteeing departures from a Gaussian profile for any given image. 

Photometry of each source passing the cuts above is performed in each aperture size, before a final cut to remove any sources that have SNR < 3 in the small aperture. At this point a detection table is constructed from all sources that pass the above cuts, which is stored in a FITS extension upon storage of the image (\cref{sec:single_storage}), as well as in the \gloss{photometry.single} table of the \kadmilos\ database.

\begin{figure*}
    \centering
    \includegraphics[width=\linewidth]{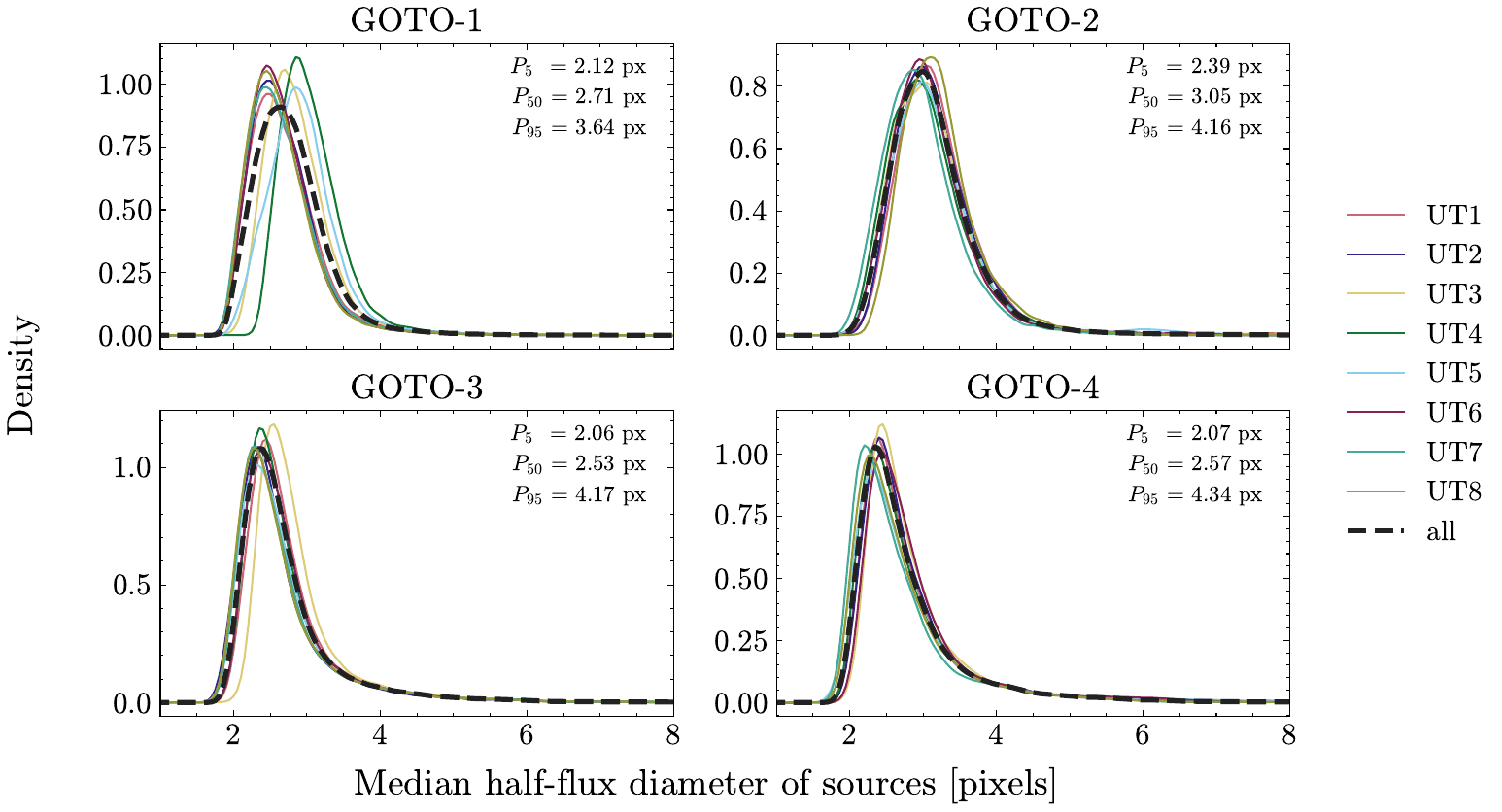}
    \caption{Density distributions of source HFD, calculated from medians in individual GOTO \gloss{single} images. Individual UT distributions are shown (thin coloured lines), along with the overall distribution per mount (thick dashed line). Density estimates were determined using a Gaussian kernel with bandwidth using the method of \citet{scott_kde}. The distributions were initially filtered to remove the low number of images with median HFD values of $>15$\,px. The more varied performance of GOTO-1 is reflective of a more aged collimation of that mount compared to the others.}
    \label{fig:goto_source_image_hfd_distribution}
\end{figure*}

 The distribution of HFDs for all sources in useful images, split by mount and UT, are shown in \cref{fig:goto_source_image_hfd_distribution}. The 50th and 95th percentiles of the HFD distribution for most UTs across the array are $P_{50}\simeq2.6$, and $P_{95}\simeq4.2$\,pixel.

\subsubsection{Astrometry}
\label{sec:single_astrometry}

Positions of high SNR (>20) sources in an image's detection table are used to determine a World Coordinate System (WCS) solution for each image, via the {\sc astrometry.net} package \citep{astrometrynet}, using their 5000-series index files, built from \textit{Gaia} DR2 \citep{gaia,gaiadr2}. With the large, distorted images from GOTO UTs, the main \texttt{solve-field} routine of {\sc astrometry.net} is called multiple times per image to improve the overall solution since it was found by experimentation to perform poorly in the outer reaches of the field of view otherwise.

First, the central quarter area of the image is solved without using any higher-order tweaks such as Simple Image Polynomials \citep[SIP;][]{sip}. Initial guesses for the pixel scale, based on the camera's configuration file (\cref{sec:dagversioning}), and the central Right Ascension (RA) and Declination, from the telescope pointing information in the FITS header of the image, are used. This initial solution provides an estimate for the footprint of the image on sky, which is used in turn to query the local copy of the ATLAS-REFCAT2 catalogue (\cref{sec:postgresql}) to find calibration sources within the image, fetching, at most, 50000 sources in descending brightness order. Cross-matching of image sources and calibration sources is performed using their RA and Declination coordinates (respectively, from the initial WCS solution of the image, and those stored in ATLAS-REFCAT2). Any not matched within a $3^{\prime\prime}$ radius, or which are ambiguously matched (i.e. not one-to-one), are discarded. A simple linear photometric calibration is then applied to search for outliers in a fitted relation between instrumental GOTO magnitudes and the ATLAS-REFCAT2 magnitudes in the filter used for later photometric calibration (\cref{sec:single_photometric_calibration}). This step effectively removes spurious bright--faint matches, which occur more frequently when the depth of a GOTO image is not comparable to the faintest calibration sources, such that the fraction of sources appearing in both the image and the calibration catalogue becomes diminished.

The image is then divided into four sub-tiles, by splitting the image in along each axis. The process above is then repeated for each sub-tile: performing a no-tweak WCS solution and cross-matching sources to ATLAS-REFCAT2. For each pass, the parent cross-matching table is updated with new cross-matches that pass the cuts for that sub-tile's WCS solution within the given sub-tile's footprint.

This parent cross-match table is then used to fit SIP corrections to the spatially-varying astrometric residuals of positions between the image and catalogue positions. These take the form of two third-order polynomial surfaces describing the residuals in each image axis. The fitting occurs iteratively, a maximum of 10 times. Each iteration: 

\begin{enumerate}
\item{Computes a SIP correction to the no-tweak WCS based on the current cross-match table.}
\item{Calculates the median astrometric residual of sources after including this SIP correction.} 
\item{Checks the solution is converging by ensuring there are more cross-matches and a smaller median residual compared to the previous iteration.}
\item{Updates the cross-match table for the next iteration.}
\end{enumerate}

The checks within (iii) are early stopping criteria meaning the iterative fitting almost always terminates before the maximum 10 iterations.

\begin{figure}
    \centering
    \includegraphics[width=\linewidth]{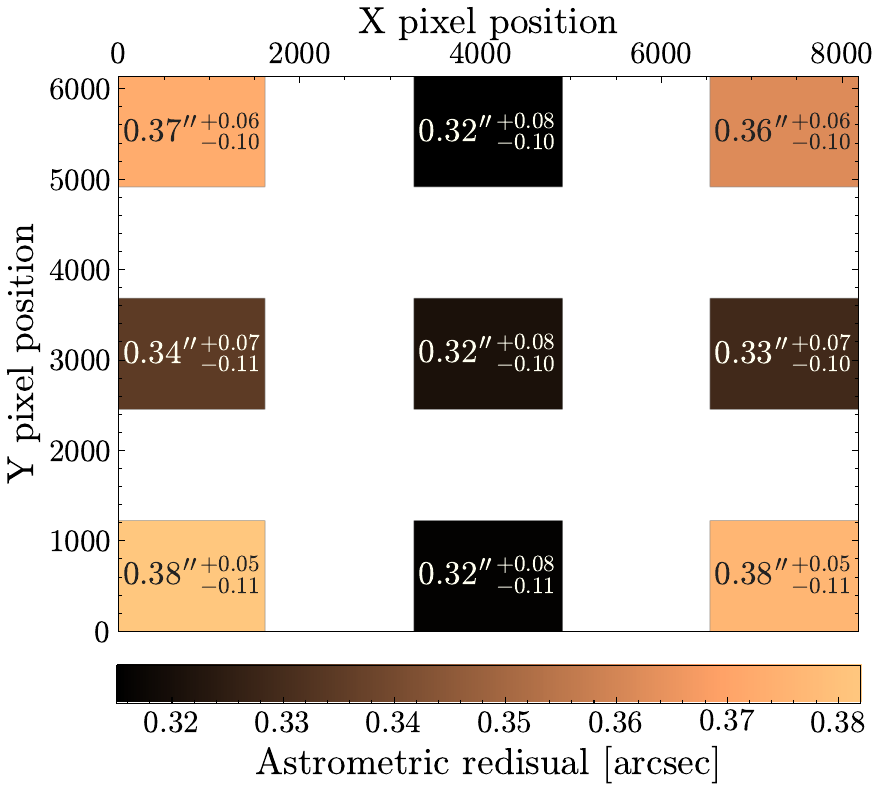}
    \caption{Heatmap of astrometric residuals of high SNR GOTO source centroids from ATLAF-REFCAT2 catalogue \textit{Gaia}-based positions. Quantities were calculated from 5000 randomly selected images and are shown with uncertainty based on the $P_{16}$ and $P_{84}$ values of the distribution. \kadmilos\ splits the images into 25 ($5\times5$) sub-tiles for astrometric stats, but only calculates for central and extrema sub-tile in each axis -- i.e. the coloured regions of this figure. For an estimate of the total GOTO astrometric residual of detections, an additional statistical component should be considered in the low SNR regime (see text).}
    \label{fig:astrometic_residual_heatmap}
\end{figure}

With the final SIP-tweaked WCS solution in hand, a final, tighter cross-matching radius of $1.5^{\prime\prime}$ is applied to create the final cross-matched table of sources. Statistics for the median and standard deviation of residuals are calculated after splitting images into $5\times5$ sub-tiles, which are then stored in the FITS header and database. A visualisation of these residuals is shown in \cref{fig:astrometic_residual_heatmap}, indicating \mbox{$0.32$--$0.38^{\prime\prime}$}residuals, depending on source position -- around a quarter to a third of a GOTO pixel. Collimation procedures try to maximise the area of ``good'' PSF coverage, which of course prioritises central areas of the image, the typically poorer PSF performance in the corners is evident from a slightly elevated astrometric uncertainty, but this is limited to a $0.05^{\prime\prime}$ change across a typical image. A more detailed view of the telescope-specific astrometric residuals can be seen in \cref{fig:astrometric_residual_violins} - the effect of different collimation performance (a maintenance task that is typically performed per telescope) is evident, although median differences are small, at the level of $\sim0.1$\,arcsec.

It should be stressed that this astrometric performance is based on the statistics of high SNR sources. An additional source of positional uncertainty, and one particularly relevant for discovery of faint infant transients, is the centroid of the source itself. One may estimate centroid uncertainty for a given SNR, assuming an idealised Gaussian profile as:

\begin{equation}
    \sigma_\mathrm{cen} \simeq \frac{\mathrm{FWHM}}{2\sqrt{2\ln 2} \cdot \mathrm{SNR}},
\end{equation}

For a marginal $\mathrm{SNR}=5$ source in a typical $\mathrm{FWHM}=4.3$\,pixel image (\cref{sec:single_source_detection,fig:goto_source_image_hfd_distribution}), $\sigma_\mathrm{cen} \simeq 0.4$\,pixel $\simeq 0.5^{\prime\prime}$. This positional uncertainty is representative of that due to the source itself, and should be considered in quadrature with the local image-level astrometric residuals to better describe the absolute sky-position uncertainty of marginal sources. We find no strong dependence of $\sigma_\mathrm{cen}$ with field density or extended structures due to the whole-sky availability of good-quality astrometric positions from \textit{Gaia} and the robust background subtraction and point-source extraction in our GOTO data.

\subsubsection{Photometric Calibration}
\label{sec:single_photometric_calibration}

The workhorse filter of GOTO is a wide optical filter, \textit{L} \citep[400--700\,nm; see][]{goto_prototype}. Although each UT houses a filter wheel and an array of filters, \textit{L} is currently used exclusively during transient discovery observations and so we only discuss its calibration here. \textit{L} broadly covers the more standard \textit{g} and \textit{r} filters used in, e.g., Sloan Digital Sky Survey \citep{sloan_photometry} and Panoramic Survey Telescope and Rapid Response System \citep[\panst;][]{panstarrs}. As such, photometric calibration is done using these comparison filters, including a colour term given the significant width of the GOTO filter.

The detection table of an image (arising from \cref{sec:single_source_detection}) is filtered for sources that have: an astrometric cross-match with ATLAS-REFCAT2 (\cref{sec:single_astrometry}), a high SNR ($>75$), and that are not within 20\,pixels of the image edge. The cross-matched catalogue photometry in ATLAS-REFCAT2 for each source is then used to further filter for those sources which have: 
\begin{enumerate}
    \item A \textit{g--r} colour within 0.0--1.3\,mag.
    \item A brightness of $11.5 < m_g < 20.5$\,mag.
    \item No nearby confusion sources, by requiring less than 10\% of their flux to be contained within $7.5^{\prime\prime}$ of their position based on \textit{Gaia}.
    \item A comparatively stable magnitude determined from the $\chi^2$/DOF value in ATLAS-REFCAT2, from \panst\ epoch photometry \citep{atlasrefcat2}.
    \item A low but non-zero proper-motion value of $<0.1^{\prime\prime}$ per year. Zero proper-motion values may be indicative of extragalactic sources.
\end{enumerate}

\begin{figure}
    \centering
    \includegraphics[width=\linewidth]{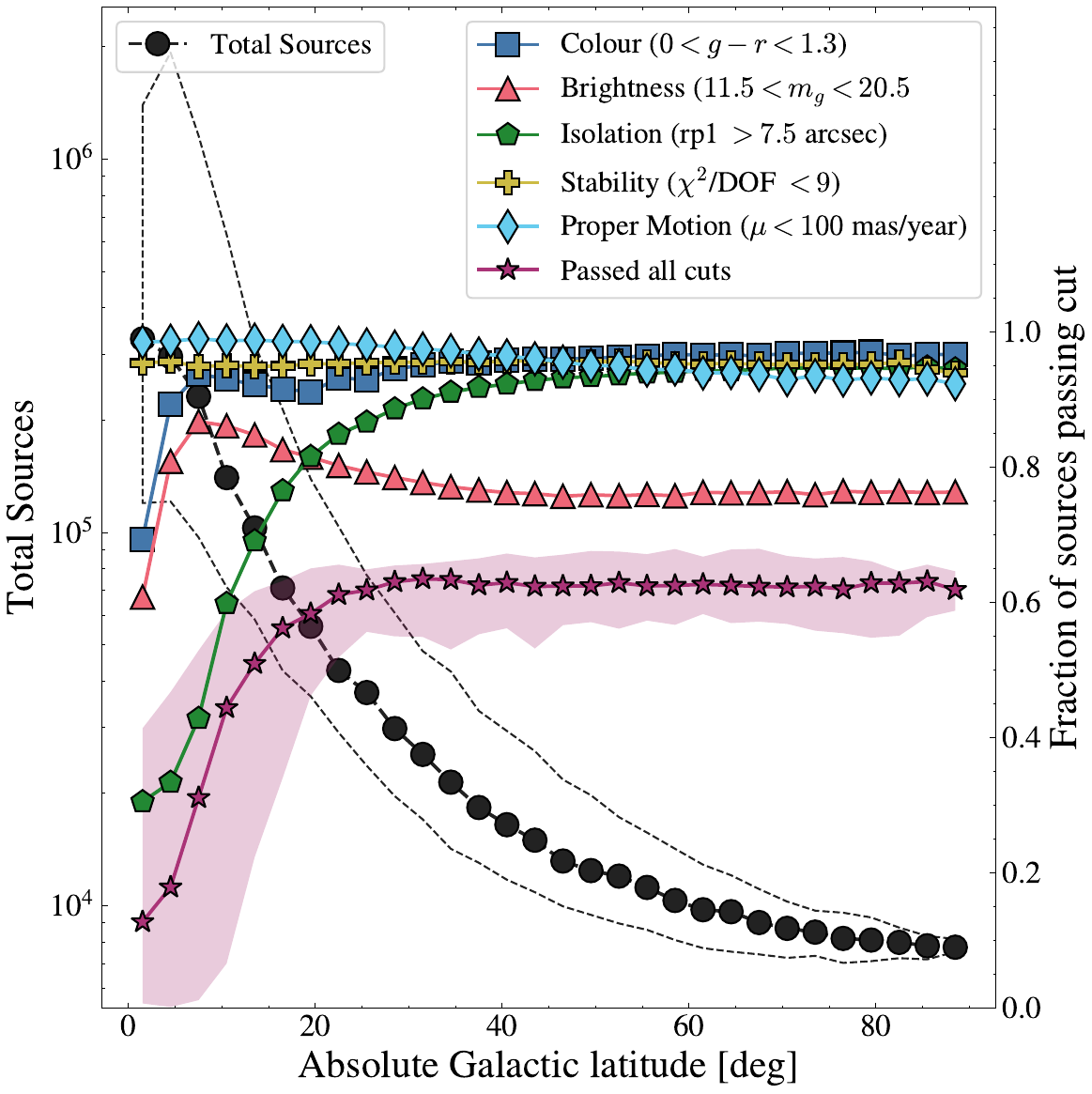}
    \caption{The number of ATLAS-REFCAT2 calibration sources in 1.4\,deg cone searches at central locations of GOTO tile positions, shown as a function of Galactic latitude (black circle markers, left axis). The thin black dashed lines indicate the $P_{5-95}$ range of these values. The right axis and differently coloured marker shapes show the fraction of those total sources that survive each photometric quality cut detailed in \cref{sec:single_photometric_calibration}, as well as those that survived all cuts. The shaded region for ``Passed all cuts'' source counts similarly indicate the $P_{5-95}$ range of values. Ranges are omitted for individual cuts for visual clarity. Survival rate of sources drops dramatically at low Galactic latitude, primarily due to a lack of isolated sources, but the survival number (total sources $\times$ passed all cuts) is relatively stable.}
    \label{fig:atlas_refcat2_cuts}
\end{figure}

A population of cone searches across the sky, designed to mimic a typical GOTO UT field of view, were used to determine the survival of ATLAS-REFCAT2 catalogue sources in the face of these criteria. The results are shown in \cref{fig:atlas_refcat2_cuts}, plotted as a function of Galactic latitude, the principal component affecting the survival of sources. Most cuts are relatively stable across the sky and have a high fraction of survivors, with colour and brightness cuts only dipping in the central few degrees of the Galactic plane, primarily due to extinction effects. Isolation, as expected, drops quite dramatically in the crowded low latitude fields, but this is offset by an increase in available sources. For the regions of primary interest for transient searches, namely $|l| > 10$\,deg, around 55-65\% of sources from the catalogue survive. In practice, the requirement for a high SNR detection in the GOTO image is typically the limiting factor. 

\begin{figure*}
    \centering
    \includegraphics[width=\linewidth]{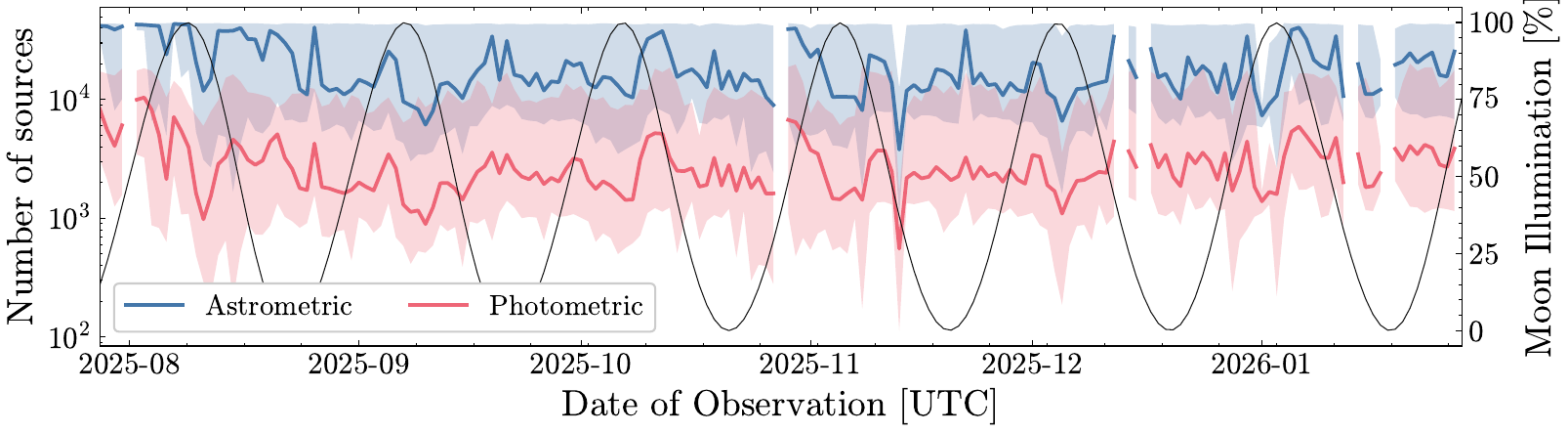}
    \caption{The number of sources used in the astrometric and subsequent photometric calibration of GOTO images. Solid lines indicate the median value (binned to daily frequency), with the shaded regions showing the $P_{5-95}$ ranges. Also plotted is the moon illumination (thin dark line, right axis), correlating with small dips in source numbers around full illumination due to an imparted lower sensitivity of GOTO images. The stable upper bound on the number of astrometric matches is a consequence of the 50000 maximum sources fetched from ATLAS-REFCAT2 per footprint.}
    \label{fig:goto_astromphotom_n_source}
\end{figure*}

In \cref{fig:goto_astromphotom_n_source} we show the number of astrometric and photometric sources used in the calibration of images. A minimum of 100 sources are required to be used in the calibration steps, in order to robustly over-constrain a spatially-varying calibration.

In order to capture the spatially-varying nature of the calibration across a UT field of view, a surface, $z = f(x,y)$ must be defined, where $x$ and $y$ are pixel coordinates of the image, and $z$ is some spatially-varying quantity of the image. This surface must be constructed from a sequence of irregularly sampled positions -- i.e. positions of stars -- rather than an idealised dense, regular mesh of points. Typical approaches use flavours of low-order polynomial or spline for $f$ (as indeed is the case for astrometric residuals, e.g. \citealt{sip}). Options such as splines \citep{splines} offer advantages through providing a smoothed representation of these intrinsically noisy and irregularly spaced tie-points. However, splines are limited in their capability to model sharply diverging areas of the surface, or very localised phenomena (without consequently over-fitting more smoothly behaving regions). A solution was found in the framework of Radial Basis Functions (RBFs), specifically using the kernel of a thin-plate (polyharmonic) spline \citep[e.g.,][]{rbf} -- an example application of RBFs for astronomical data calibration can be found in \citet{mitchell2025}. The surface, $z$, is produced through a linear combination of RBFs, each produced from the measurement and position of a star that passed the above filtering. To ease computational cost, a thinning method is applied to these stars which subsamples those in dense regions, whilst retaining isolated ones, thereby maintaining information on the spatial behaviour with fewer measurements. Practically, the number of neighbours ($N$) within a certain pixel radius ($r$) is calculated for each star. The probability of retaining a star $P_\mathrm{ret}$ is then:

\begin{equation}
    P_\mathrm{ret} = \begin{cases}
1, & \text{if } N < N_\mathrm{lim} \\
N_\mathrm{lim}/N, & \text{otherwise},
\end{cases}
\end{equation}

where $N_\mathrm{lim}$ is a limit on the number of close neighbours a star can have before sub-sampling is enacted. In \kadmilos, $N_\mathrm{lim} = 10$ and $r = 180$\,pixels are used.

\begin{figure}
    \centering
    \includegraphics[width=\linewidth]{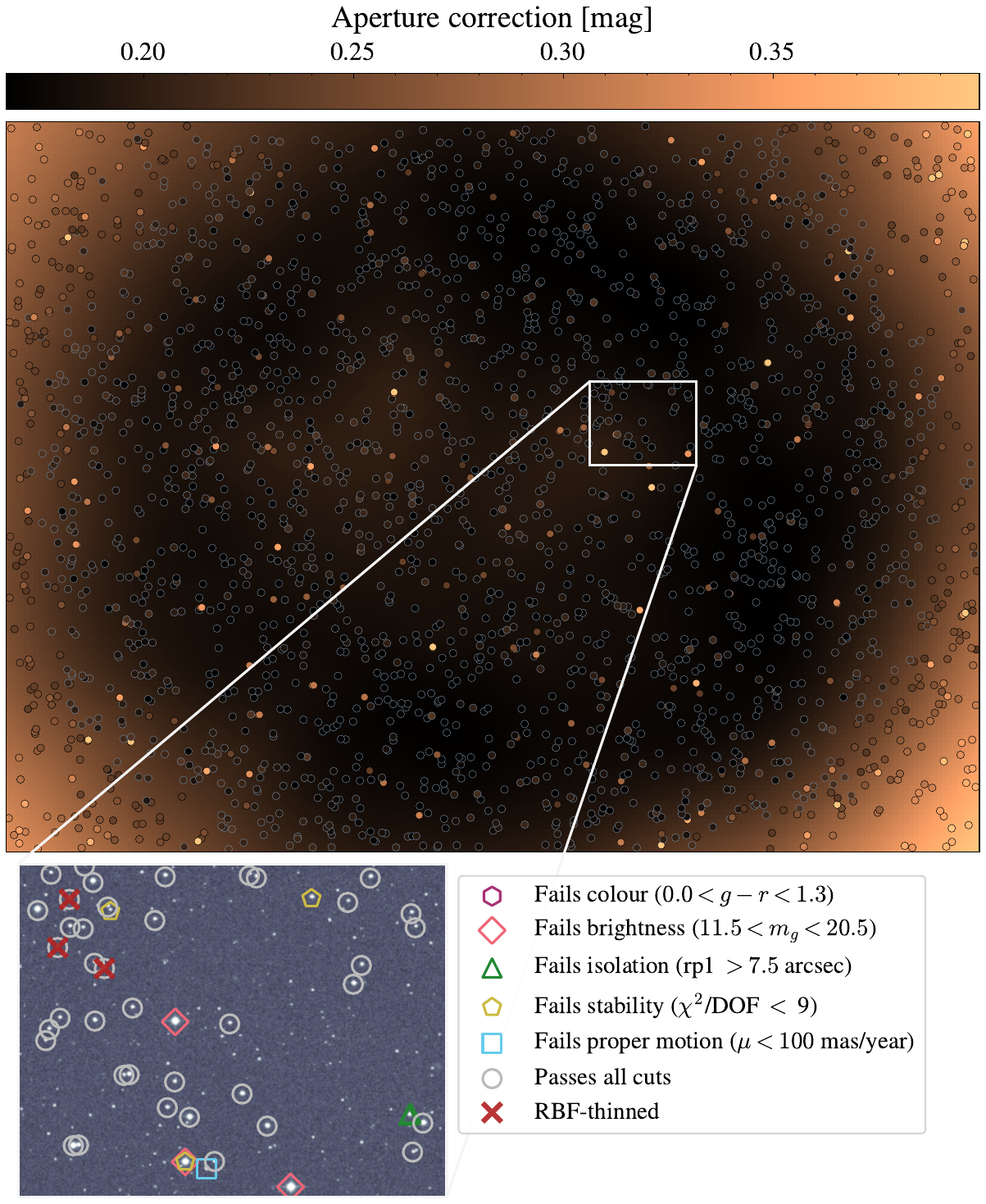}
    \caption{A heat map of the aperture correction Radial Basis Function (RBF) surface for a representative GOTO-1 UT-1 image. Overlaid on this heat map are scatter points of detections used to construct the surface. They are colour coded on the same scale and outlined for visual clarity. A zoom-in region is indicated, showing the underlying image data with coloured markers surrounding high SNR detections that fail the various criteria (i.e. low-SNR sources are already excluded and not marked in the figure). The criteria are described further in \cref{fig:atlas_refcat2_cuts,sec:single_photometric_calibration}. Sources may be marked multiple times, indicating they failed multiple criteria. No detections failed the colour cut in this zoom-in region.  Further, those culled due to the probabilistic thinning technique (see text) are overlaid with crosses. Only detections passing all cuts and not thinned are used in the construction of the RBF surface. A comparison of the zoom-in detections with the outlying scatter points in the aperture correction heat map shows they are due to stars with close neighbours. Although the \mbox{\texttt{rp > $7.5^{\prime\prime}$}} criterion limits the impact of egregiously crowded sources, neighbours can still affect aperture photometry of those making it through the cut. Nevertheless, the heat map shows that these individual outliers do not influence the surface, which maintains a good description of the overall image quality, and demonstrates the robustness of using RBFs in this domain.}
    \label{fig:apcor_rbf}
\end{figure}

An RBF surface is constructed to describe the aperture correction of the image (i.e. the difference in magnitudes between the small and large aperture photometry (\cref{sec:single_source_detection}), using all detections that pass the above criteria. This surface is then applied to the small aperture photometry of all stars to get their large-aperture-corrected magnitudes. A visualisation of the detection filtering and construction of this aperture correction surface is shown in \cref{fig:apcor_rbf}. The shape of this surface highlights the collimation challenge with such fast aperture telescope design that utilises almost the entire unvignetted field of view. The area of consistent performance is maximised during the collimation process, resulting in a ring of best performance with a very slight degradation at the centre, and the unavoidable degradation in the corners.

A second-order polynomial is then fitted to the colours $g-r$ and $g-L_\mathrm{inst}$ using all filtered sources in the image, where $L_\mathrm{inst}$ is the instrumental GOTO $L$-band magnitude, aperture-corrected using the RBF surface calculated above. The relation is used to calculate residuals of stars -- the difference in magnitudes between what is measured and that predicted from the polynomial relation. These residuals are used as tie points for another RBF surface calculated analogously to the aperture correction, which defines a delta zeropoint surface, allowing the calibration to account for differential atmospheric transparency, for example, across the image.

\begin{figure*}
    \centering
    \includegraphics[width=\linewidth]{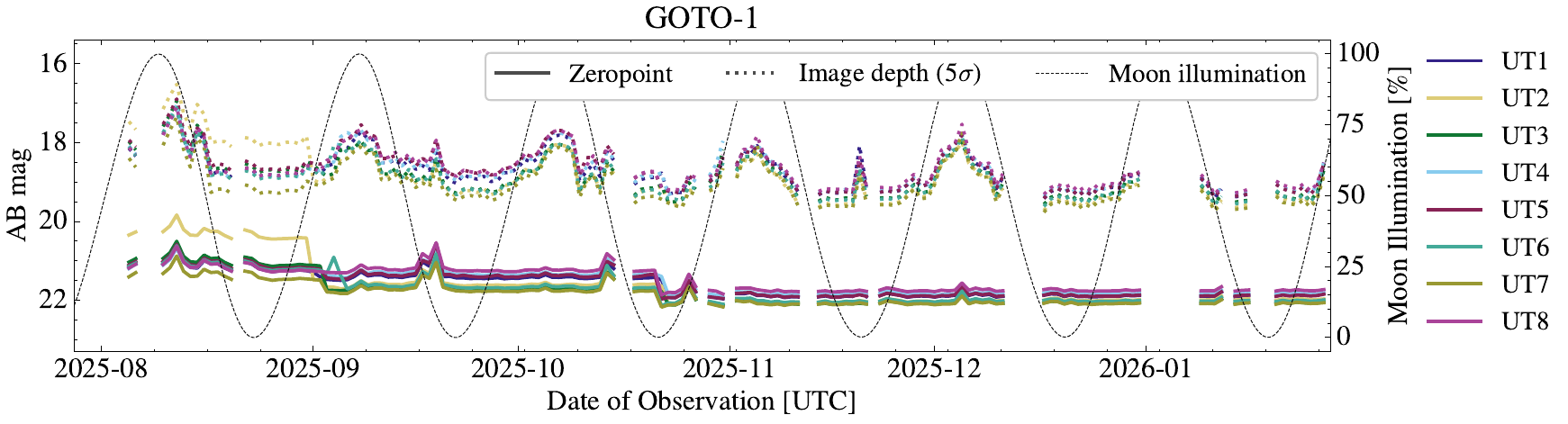}
    \caption{Results of photometric calibration of 45\,second \gloss{single} GOTO-1 images, split by UT (colours). Shown on the plot are daily bins of the zero-point (thick solid lines) and median detection magnitude of $5\sigma$ sources in the images (thick dotted lines). Overlaid is the moon illumination (thin dashed line), causing a reduction in image depths around full moon. Periods of no data are either due to bad weather, technical downtime, or nights when no 45\,second exposures were taken (e.g. during intensive external trigger follow-up, which employs a different observing strategy.) The effect of maintenance trips to clean the primary and secondary mirrors can be seen as step changes in the zeropoint performance -- late Aug 2025 for UT2 and mid Oct 2025 for the remaining UTs.}
    \label{fig:zeropoints_1}
\end{figure*}

Results of the calibration in the form of zeropoints for GOTO-1 UTs are shown in \cref{fig:zeropoints_1}, with equivalent plots for remaining mounts given in \cref{fig:zeropoints_other}. Most UTs had shown overall little degradation compared to the initial installation and design sensitivity \citep{dyer_thesis}. However, GOTO-1 UT2 degraded noticeably faster than other UTs (which are all relatively consistent). A more rapid build up of dust on the secondary mirror of this UT, for unknown reasons, was found. A maintenance trip in September 2025 managed to recover close to design sensitivity of this UT following an intensive cleaning operation. Subsequent maintenance trips to La Palma (GOTO-1 and GOTO-2) and Siding Spring (GOTO-3 and GOTO-4) in late Oct and late Nov 2025, respectively, are evident by increases in the zeropoint values of UTs (\cref{fig:zeropoints_1,fig:zeropoints_other}). A more detailed discussion of the zeropoint behaviour as related to the telescope hardware will appear in Steeghs et al. (in prep). Short spikes of lowered sensitivity related to poorer ambient conditions, with the overall sky-brightness effect of the Moon evident in \cref{fig:zeropoints_1} and \cref{fig:zeropoints_other} given the blue-optical wavelength coverage of the $L$-band. We emphasise that the performance shown in these figures relates to 45\,second \gloss{single} images of GOTO. In practice, both the sky-survey and follow-up modes employ stacking of \gloss{single} images into \gloss{set} images to improve sensitivity (\cref{sec:set_image_generation,fig:set_image_depths}).

\subsubsection{Storage format of \gloss{single} images}
\label{sec:single_storage}

After the processing of previous subsections, the \gloss{single} image data are normalised to one second by dividing the pixel values by the exposure time. \gloss{single} images are then stored as a multi-extension FITS file with the following extensions:

\begin{center}
\begin{tabular}{ p{2.0cm}  p{5.4cm} }
 Extension & Description \\ 
 \hline
 science & Reduced, cleaned, background-subtracted, normalised \gloss{single} image data. \\  
 science\_mask & Bit-wise integer mask of bad or otherwise affected pixels. \\
 science\_bkg\_rms & Spatially-varying RMS map determined from background fitting. \\
 science\_phot & Source position and photometry table, including cross-matched catalogue data. \\
\end{tabular}
\end{center}

The location of the FITS file, as well as statistics on the image quality (HFD, number of detections), astrometric and photometric fits is stored as a record in the appropriate \postgres\ table. This record holds identifiers of the calibration frames (\gloss{bias} etc.) used in the processing of the \gloss{single} image. The two RBF objects generated for each image during photometric calibration (\cref{sec:single_photometric_calibration}) are additionally serialised to bytes using {\sc dill}\footnote{\url{https://dill.readthedocs.io/en/latest/}} and stored directly within the \postgres\ record for the image. The (de)serialising of these objects is a quick process, and provides a space efficient means to store them. This allows the surfaces to quickly be loaded and used during subsequent analysis (e.g. forced photometry, \cref{sec:marshall_forcedphotometry}).

\subsection{Creation and science processing of \gloss{set} images}
\label{sec:set_image_generation}

As detailed in \cref{sec:goto_observing_strategy}, each GOTO pointing consists of one or more \gloss{single} images meant to be combined into a stacked \gloss{set} image. The addition of read-noise (cf. a single deep exposure) is offset by mitigation of cosmic-ray artefacts, and the rejection of short-lived poor conditions or technical issues. 

The \gloss{single} images to be combined into \gloss{set} images are tracked via unique combinations of UT, mount and `setnum' values -- incrementing integer designating each set. Once all \gloss{single} images from a given set have been processed, the workflow to combine them is initiated.

\subsubsection{Determination of relative image weights}
\label{sec:relative_image_weights}

The combination of \gloss{single} images is done in a weighted fashion, closely following the formalism of \citet{sdss_combine_weights} to generate relative weights for each image. Namely:

\begin{equation}
    w_i = \frac{T_i}{\mathrm{FWHM}_i^2},
\end{equation}

where $w_i$ is the weight of the $i$th image, and $T_i$ and $\mathrm{FWHM}_i^2$ are relative values of the transparency (based on zeropoint) and FWHM of the image, with respect to the `best' image in the set. The best image in the set then naturally has $w=1$. Any image with $w<0.6$ is excluded from the combination process as it was found these are typically compromised images, due to various issues such as mount tracking errors, scattered light from artificial sources, rapid degradation of ambient conditions, or similar anomalous phenomena. In the final \gloss{set} images they contribute largely noise, having minimal, sometimes negative, influence on the quality of the stacked image, even in spite of their lower weight.

\subsubsection{Astrometric registration}
\label{sec:set_astrometric_registration}

Between the individual exposures of a pointing, the GOTO mounts dither their position by several arcsec, which can depend on the number of sub-exposures although a 4-point pattern is typical for both sky-survey and follow-up observations discussed here. This approach shifts bad pixels spatially, such that stacked pixels are almost always composed of mostly good pixels. The exception to this can be large CCD defects or defects in the optics (e.g. mirror blemishes). The dithering step introduces the requirement that the processed \gloss{single} images in a set are astrometrically aligned prior to combination. (Such alignment would most likely be required even in the absence of dithering due to, e.g., imperfect tracking, shifting airmass).

\spalipy\ \citep{spalipy} is an astronomical-image alignment tool utilising source detection centroids to calculate a differential transformation rather than alignment to an absolute frame such as a WCS (see \cref{app:spalipy} for more details). To transform the small shifts between \gloss{single} images of a set, an affine transformation ignoring shear terms proves sufficient. This affine transformation is determined using several hundred to several thousand source positions. The first \gloss{single} in the set is used to define the pixel coordinate system, and others are aligned to this, including their background RMS and bad pixel maps. Any failed alignments are removed from the list of \gloss{single} images to combine.

\subsubsection{Weighted combination}
\label{sec:set_weighted_combination}

Alongside the overall relative image weights (\cref{sec:relative_image_weights}), the combination of the \gloss{single} images employs a further pixel-wise weighting. This weighting is the inverse variance of the background, calculated from the background RMS map stored within the \gloss{single} FITS file (\cref{sec:single_storage}).
Pixel-wise combination is done using a weighted mean, with sigma-outlier clipping. Due to memory demands of holding 10-12 GOTO images (in the case of deep observations) in memory, plus their background maps, per \gloss{set} image generation, this stage is performed in sections. The stack is split into 100 chunks along the longer image axis, and the data are held in memory-mapped files on fast, local disks via {\sc numpy} \citep{numpy}. The combined data array generates the \gloss{set} image.

\subsubsection{Science processing}
\label{sec:set_science_processing}

The science processing of \gloss{set} images follows that of the \gloss{single} images, as detailed in \cref{sec:single_source_detection,sec:single_astrometry,sec:single_photometric_calibration}. Although the astrometric calibration may be largely redundant, the benefit of additional depth and suppressed noise in the \gloss{set} image means a recalibration ensures the stored astrometric solution most accurately reflects the final \gloss{set} image.

\begin{figure*}
    \centering
    \includegraphics[width=\linewidth]{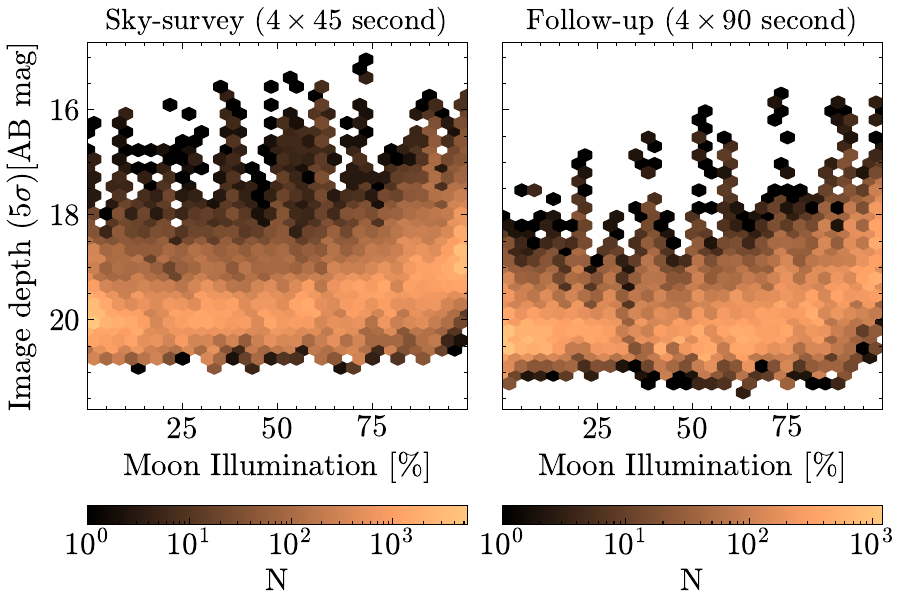}
    \caption{2D histograms of GOTO \gloss{set} image depths as a function of moon illumination in the $L$-band. The two main observing strategies are indicated as titles above each subplot. Note that the colour scaling of bins is logarithmic.}
    \label{fig:set_image_depths}
\end{figure*}

The image depths of \gloss{set} images taken in the regular sky-survey mode ($4\times45$\,second) and triggered follow-up of external alerts ($4\times90$\,second) are shown in \cref{fig:set_image_depths}. Sky-survey observations deliver a typical $5\sigma$ depth of $m_L \sim 20$\,mag, reducing in sensitivity by $\sim$1\,mag in the brightest lunar conditions ($\gtrsim85\%$ illumination). Follow-up of external triggers delivers $m_L \sim20.5$\,mag, approaching $\sim21$\,mag in better conditions (and of course likewise suffering from the same lunar impact as the sky-survey).

\subsubsection{Storage format of \gloss{set} images}
\label{sec:set_storage_format}

Although \gloss{set} images are stored following processing, they are typically not interacted with by end users, and are picked up immediately by the next stage of the pipeline, where they are written with the full suite of data and table extensions. The reader is directed to \cref{sec:difference_storage_format} for information on this format.

\subsection{Difference image analysis}
\label{sec:DIA}

Difference image analysis (DIA) is a common technique in transient sky surveys, whereby a historical, high-quality image of the sky (ideally taken with the same telescope and optical set-up) is used to subtract constant-flux sources, leaving behind only transient, variable, and moving sources \citep{dia}. These historical images are referred to as \gloss{template} images, with the result being a \gloss{difference} image (\cref{tab:nomenclature}).

The observation strategy for \gloss{template} images is provided in \cref{sec:goto_observing_strategy}. The noise of the \gloss{set} and \gloss{template} are, to first order, added in quadrature during DIA to give the overall noise level of the \gloss{difference} image. Using a much deeper \gloss{template} image ensures only a small noise contribution to the \gloss{difference} image, which consequently retains a depth comparable to the original \gloss{set} image.

A process of manual vetting of \gloss{templates} was done using consortium members via a private Zooniverse\footnote{\url{https://zooniverse.org}} platform. After each vetting they were incrementally ingested into a specific \postgres\ table, where they could be searched for by the DIA stage of the low-latency processing DAG \singlesetcreation. 

\subsubsection{Creation of \gloss{difference} images}
\label{sec:difference_image_creation}

A \gloss{template} search is initially made to find one matching the mount, UT, filter and \gloss{tile} of an incoming \gloss{set} image. However, it was found that, visually, the impact from using a \gloss{template} image acquired on one mount at a given site to perform DIA of a \gloss{set} image acquired on the other mount was not significantly degrading the quality of the subtraction. Rather, issues with DIA mainly arise due to data quality issues, including poor conditions. Therefore, to streamline the generation of \gloss{template} images, only one set of valid \gloss{template} images were obtained per site (rather than per mount). As such, if a \gloss{template} is not initially found, the requirement for having the same mount is then relaxed (although originating from the same site remains enforced). A final check in the search for a valid template is to ensure the distance between the centre of the \gloss{set} and \gloss{template} images is $\leq0.35$\,degrees. Due to distortions in GOTO images in the outer regions of the FoV, severely mismatched images on-sky prove difficult to accurately align. Due to the similar footprint across mounts, and the consistent \gloss{tile} grid used, this limit is violated only infrequently.

The on-sky overlap between the \gloss{set} and \gloss{template} is calculated using their respective astrometric footprints in order to trim the source detection catalogues for each image of those falling outside the overlap. These overlapping sources are then used to align the \gloss{template} to the pixel-coordinate system of the \gloss{set} image, including its background RMS map. As for the registration of the constituent \gloss{single} images (\cref{sec:set_astrometric_registration}), \spalipy\ is used to calculate and perform the alignment. In this scenario, however, complex transformations can be required, and so the full transformation capability of \spalipy\ is used, to allow (third order) spline corrections to additionally warp the affine transformation to account for non-linear distortion of the images.

With the \gloss{template} data and background RMS map registered to the \gloss{set} image, the image differencing is performed using a multi-threaded version of the High Order Transform of Psf ANd Template Subtraction code \citep[HOTPANTS;][]{hotpants}\footnote{\url{https://github.com/Lyalpha/hotpants}}. The algorithm of HOTPANTS is itself building upon the work of \citet{isis1} and \citet{isis2}, through the use of a spatially-varying regulated kernel which, when convolved with the \gloss{template} image, will generate a new image with a PSF and scaling matching (as closely as possible) that of the \gloss{set} image. Prior to running the subtraction, the masks of the \gloss{template} and \gloss{set} images are logical-or combined, to build a mask for the difference image. Owing to sub-pixel shifts in alignment, the \gloss{template} mask would nominally suffer from interpolation artefacts and incorrect propagation of its integer mask. To circumvent this, we apply a `nearest' interpolation technique, which preserves the values of the initial mask, at the small expense of missing some partial pixel overlaps.

\subsubsection{Analysis of \gloss{difference} images}
\label{sec:difference_image_analysis}

The analysis of the \gloss{difference} image data begins with source detection, as is done for both \gloss{single} and \gloss{set} images (\cref{sec:single_source_detection}). Unlike in those latter cases, additional checks are here put in place to test the validity of detections owing to the potential for DIA artefacts and correlated noise spikes producing spurious detections in \gloss{difference} images generated via convolution \citep{isis1,isis2}. Firstly, forced photometry is applied to the original \gloss{set} image at the position of the difference candidate detections. Any candidate with a $\mathrm{SNR} < 5$ in the \gloss{set} image is discarded. A further check is made for those detections found in DIA of \gloss{set} images that themselves are composed of a low number ($<5$) of \gloss{single} images (this is the usual case). Although the combination stage (\cref{sec:set_weighted_combination}) employs pixel-wise sigma-clipping, artefacts present in just one or two images (e.g. cosmic rays, satellite trails) can still trickle through to the stacked pixel data, causing a spurious `detection' in the \gloss{difference} data. As such, the constituent \gloss{single} images also have forced photometry performed at the locations of difference candidate detections. For each source, the fraction of \gloss{single} images in which there is a $\mathrm{SNR} > 2.5$ excess of flux, $f_\mathrm{single}$, at the position of the candidate is recorded.

As the DIA stage aligns and scales the \gloss{template} image to match the \gloss{set} image, we can use the photometric calibration of the \gloss{set} image as a valid solution for the subsequent \gloss{difference} image. Due to aforementioned DIA artefacts, source extraction on subtracted data typically finds many more astrophysically-bogus sources of flux than genuine, real detections. Due to the volume of detections, it is unfeasible to expect manual inspection of even a sizeable fraction. For this reason schemes to separate `real' and `bogus' detections have been developed, typically appealing to machine-learning models to provide candidate detections with a `realbogus' score ranging from zero to one. These model implementations principally rely on high-level features extracted from the detection  \citep[e.g.][]{bloom2012}, and/or the pixel data directly \citep[e.g.][]{duev2019}. The model used by GOTO is fully described in \citet{killestein2021} and consists of a convolutional neural-network trained on GOTO pixel data using algorithmically-constructed training data, supplemented by synthetic-transient insertions. The trained model is loaded by \kadmilos\ and used for inference on each candidate detection found in the \gloss{difference} image. The previously calculated $f_\mathrm{single}$ is now used as a modulator of the native realbogus score provided by the model. Strong, spreading cosmic-ray hits in an individual \gloss{single} image, which are not fully cleaned (\cref{sec:single_cosmic_ray_bad_pixel_correction}), can particularly mimic faint sources in the ultimate \gloss{difference} image. These are particularly tricky for a model to distinguish, but show up clearly as low $f_\mathrm{single}$ values. This modulation of the model score means one-off artefacts now only propagate to the later vetting stages when they unfortunately align with a pre-existing astrophysical object, however visualisation of the \gloss{single} data during vetting makes their removal trivial. More persistent artefacts -- namely residuals around bright sources -- remain a challenge to fully clean at the model score stage, and so are the chief contaminant in later stages.

\subsubsection{Storage format of \gloss{difference} images}
\label{sec:difference_storage_format}

Data from the \gloss{set}, \gloss{template} and \gloss{difference} images used during DIA are collated and stored in a single multi-extension FITS file with the following extensions:

\begin{center}
\begin{tabular}{ p{3.2cm}  p{4.5cm} }
 Extension & Description \\ 
 \hline
 science & Reduced, cleaned, background-subtracted, normalised \gloss{set} image data. \\  
 science\_mask & Bit-wise integer mask of bad or otherwise affected pixels. \\
 science\_background\_rms & Spatially-varying RMS map determined from background fitting. \\
 science\_photometry & Source position and photometry table, including cross-matched catalogue data. \\
 template & Aligned \gloss{template} image data. \\
 template\_mask & Bit-wise integer mask associated with the \gloss{template} image. \\
 template\_background\_rms & RMS background map for the \gloss{template} image. \\
 difference & \gloss{difference} image (science - template) data. \\
 difference\_mask & Bit-wise integer mask for the \gloss{difference} image. \\
 difference\_background\_rms & RMS background map for the \gloss{difference} image. \\
 difference\_photometry & Photometry table of candidates identified in the \gloss{difference} image. \\
\end{tabular}
\end{center}

As there is no further processing required of the data, these FITS files are stored using lossy compression of the float data in image extensions to save on storage space \citep{lossycompression}. The \texttt{fpack}\footnote{\url{https://heasarc.gsfc.nasa.gov/docs/software/fitsio/fpack/fpack.html}} tiled-compression package is used \citet{fpack} with the RICE compression algorithm. The quantization for compression is chosen sufficiently high (512) to have an unnoticeable effect on the precision of photometry required ($\ll0.01$\,mag) during immediate future use such as forced photometry.

In addition, a record is stored in the \kadmilos\ database detailing the path to this FITS file, as well as high-level quantities related to the image such as number of difference image detections. Records are linked to the \gloss{set} and \gloss{template} tables, allowing one to determine the provenance of the image, back to the raw and calibration data used.

\subsection{Pipeline timings}

\begin{figure}
    \centering
    \includegraphics[width=\linewidth]{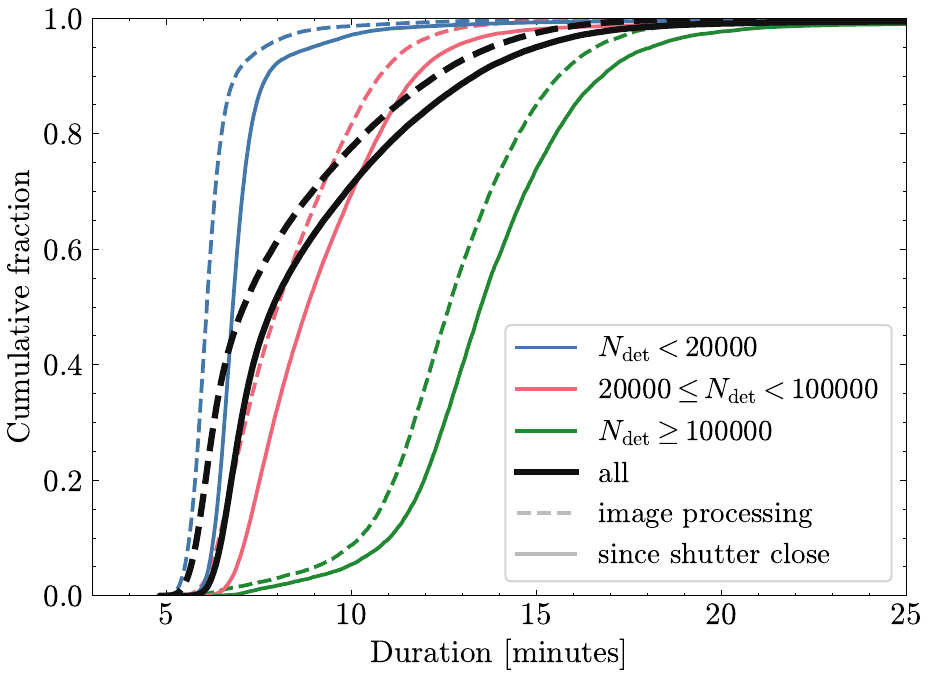}
    \caption{Total times for GOTO images to be created and analysed by \kadmilos, based on DAG run timings in Airflow. Dashed lines indicate time taken for image processing only -- i.e. the creation and analysis of a sequence of \gloss{single}, \gloss{set} and \gloss{difference} images. Solid lines measure this processing time from shutter close of the telescope for the image, thereby also including raw data readout, transfer, and ingestion stages. The overall distribution is shown in thick black lines. These are also split by the number of source detections ($N_\mathrm{det}$) in the image, since this is a principal variable affecting processing time. Note the plot has been cut at 25\,minutes and a small fraction of timings extend significantly beyond this (see text).}
    \label{fig:single_and_set_timings}
\end{figure}

The Airflow metadata database provides a means to interrogate the performance of processing by \kadmilos. Of principal interest is the time taken to complete the \singlesetcreation\ DAG (\cref{sec:dagdetails}) -- the workflow responsible for creating and analysing the various science image products. This DAG always produces \gloss{single} images, but may also produce \gloss{set} and \gloss{difference} images if the exposure being processed is the last of a pointing. The distribution of process timings for the latter case is shown in \cref{fig:single_and_set_timings}. The distribution is split on $N_\mathrm{det}$, the number of source detections found in the \gloss{single} image being processed. This quantity is chiefly responsible for the spread of timings in the distribution owing to many per-source calculations (e.g. shape, photometry) being performed, whereas image-level calculations (e.g. CCD reduction, alignment) remain constant. A small but extended tail is present in the distributions, reaching to many days or weeks. The culprits for such delays are typically issues with the original processing (e.g. a new camera was installed but did not yet have a valid \gloss{super} \gloss{flat} calibration image), hardware/network failures (e.g. database cluster being inaccessible), and delays/errors due to software mishaps. \cref{fig:single_and_set_timings} also shows the same distributions but counting the start time as time of shutter close for the telescope, better representing the entire process from writing the initial FITS file of the raw data at the respective site and ending with transient candidate detections stored in a \postgres\ table. It is important to note that these distributions do not include those images which failed processing for any reason.

Although the results in \cref{fig:single_and_set_timings} indicate images with $N_\mathrm{det} \geq 100000$ take significantly longer, these are not typical cases (cf. the `all' distribution), and, further, are of the least interest for extragalactic transients searches since they are crowded, extinction-heavy Galactic plane pointings.

\subsection{Pipeline data volumes}

\begin{figure*}
    \centering
    \includegraphics[width=\linewidth]{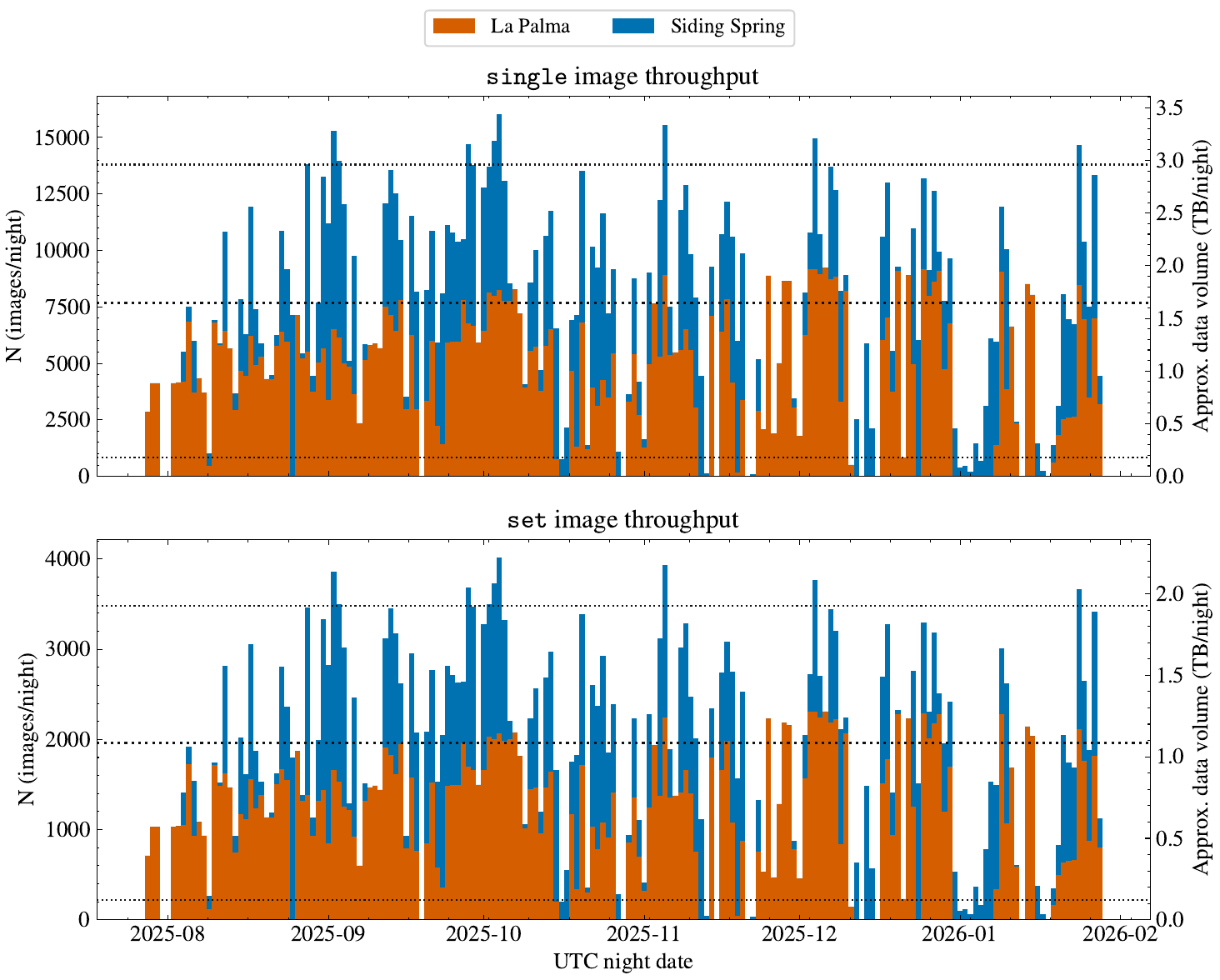}
    \caption{The nightly observed \gloss{single} (top) and \gloss{set} (bottom) image counts and approximate data volumes across both GOTO sites and over a representative date range. The sites are almost antipodal, so the data processing occurs sequentially for each site, meaning stacked counts are effectively 24\,hour windows of data processing. Horizontal bars on each panel show the 5th, 50th (median), and 95th percentile of the totals.}
    \label{fig:single_and_set_nightly_counts}
\end{figure*}

As mentioned in \cref{sec:hardware_implementation}, \kadmilos\ runs in low-latency on locally-held data in Warwick once the data are transferred from the respective sites (\cref{sec:rawtransfer}). We show in \cref{fig:single_and_set_nightly_counts} the nightly counts of \gloss{single} and their composite \gloss{set} images needing to be generated based on \textit{observed} \gloss{raw} data frames taken at each site. Image quality issues mean some frames may fail some processing stages, and so the final data products produced may be slightly below these maximal counts. Shown alongside counts are approximate data volumes assuming a typical 225 and 580\,MB file-size on disk for \gloss{single} and differenced \gloss{set} images, respectively. The median (95th percentile) disk storage volume required is $\sim2.7\,(4.9)$\,TB per 24\,hours for the stored FITS data products. Owing to the availability of filesystem block-level compression on the storage devices, which is able to further compress non-data FITS extensions, this is a small ($\sim5-10$\,percent) overestimate. The median (95th percentile) of \gloss{raw} frames observed per 24\,hours of $\sim7700 (13800)$ translates to an input data rate of $\sim0.3\,(0.6)$\,TB.

\section{Marshall implementation}
\label{sec:marshall}

A common means for visualising data from transient sky-surveys is the concept of a `marshall' system \citep[e.g.][]{ptf_marshall,atlas_transientserver}, also referred to as a Target and Observation Manager \citep[TOM; e.g.][]{tomtoolkit}. This is typically a web-interface that allows collaboration members access visualisations of survey outputs, and, when occasion demands, initiate further actions. Tools for target identification and prioritisation enable the reporting of findings promptly to the community and/or actioning of follow-up observations for science characterisation. 

The role of a marshall is typically also to perform several other important tasks, including source association (linking new detections with historical ones at the same location) and generating a wide range of high-level metadata for the source and its detections. This metadata can include contextual information from cross-matching with astronomical catalogues, calculation of macroscopic light curve properties, and application of algorithmic or machine-learned science classification \citep[e.g.][]{rapid,fleet,needle} or prioritisation \citep[e.g.][]{atlas_vra} models. Initially a per-survey or -facility component to the transient science workflow, recent implementations concentrate on aggregating source management to and from various surveys and follow-up facilities in one common interface \citep[e.g.][]{growthmarshall}.

An important further component is the development of ``broker'' systems \citep[e.g.][]{ampel_broker,fink_broker,antares_broker,alerce_broker,lasair_broker}, which take the heavy lifting of parsing data streams from multiple surveys and providing a consistent API to access and process the streams. Alongside built-in high-level classification and data-tagging tasks \citep[e.g.][]{alerce_lc_classifier,alerce_stamp_classifier}, the brokers provide entry points for contributions by the community to develop tools that add highly-specialised metadata or scores to candidates as they are processed \citep[e.g.][]{needle}. These can help de-duplicate effort in self-contained marshall systems, although typically the models are specialised and not able to be effortlessly applied to different survey data -- significant additional training of models is required at the least.

Science cases may demand some combination or subset of these architectures. For example, a spectroscopic follow-up collaboration may deploy a marshall system to ingest from several surveys -- typically via a broker of choice -- and coordinate their follow-up campaigns \citep[e.g.][]{pessto}. In contrast, survey teams themselves may have privileged additional information above that released via a forced photometry service or broker alert stream, and wish to conduct an in-house marshall in addition, in order to further facilitate their own science goals.

In the following subsections we describe the marshall developed for GOTO, from ingestion of new difference image candidate detections, to reporting of transients and triggering of external follow-up.

\subsection{Software overview}

The GOTO marshall is built using the \django\ web-framework.\footnote{\url{www.djangoproject.com}} Asynchronous task handling is done through the use of a dedicated Celery worker, similar to our Airflow implementation (\cref{sec:airflow}). The back and front-ends of the framework interact with a dedicated \postgres\ database to retrieve and update all source information.

\subsection{Ingestion of candidate detections}
\label{sec:marshall_ingest}

The processed database of \kadmilos, specifically the \gloss{difference} image records (\cref{sec:difference_storage_format}), is regularly polled by the GOTO marshall. Any image records found that are not already held by the marshall database are marked for processing. The original reason for observing the image is held by the \kadmilos\ database, and this dictates the thresholds for pulling detections associated with the image. Regular sky-survey observations require a realbogus score of $>0.7$ for detections to be ingested into the marshall, whereas observations taken due to an \gloss{event} (i.e. GW, GRB, or neutrino follow-up) have this threshold relaxed to $>0.4$, since completeness over purity is desired for these cases. Records for each difference detection passing the cut for the new image are similarly fetched and marked for ingestion.

\begin{figure*}
    \centering
    \includegraphics[width=\linewidth]{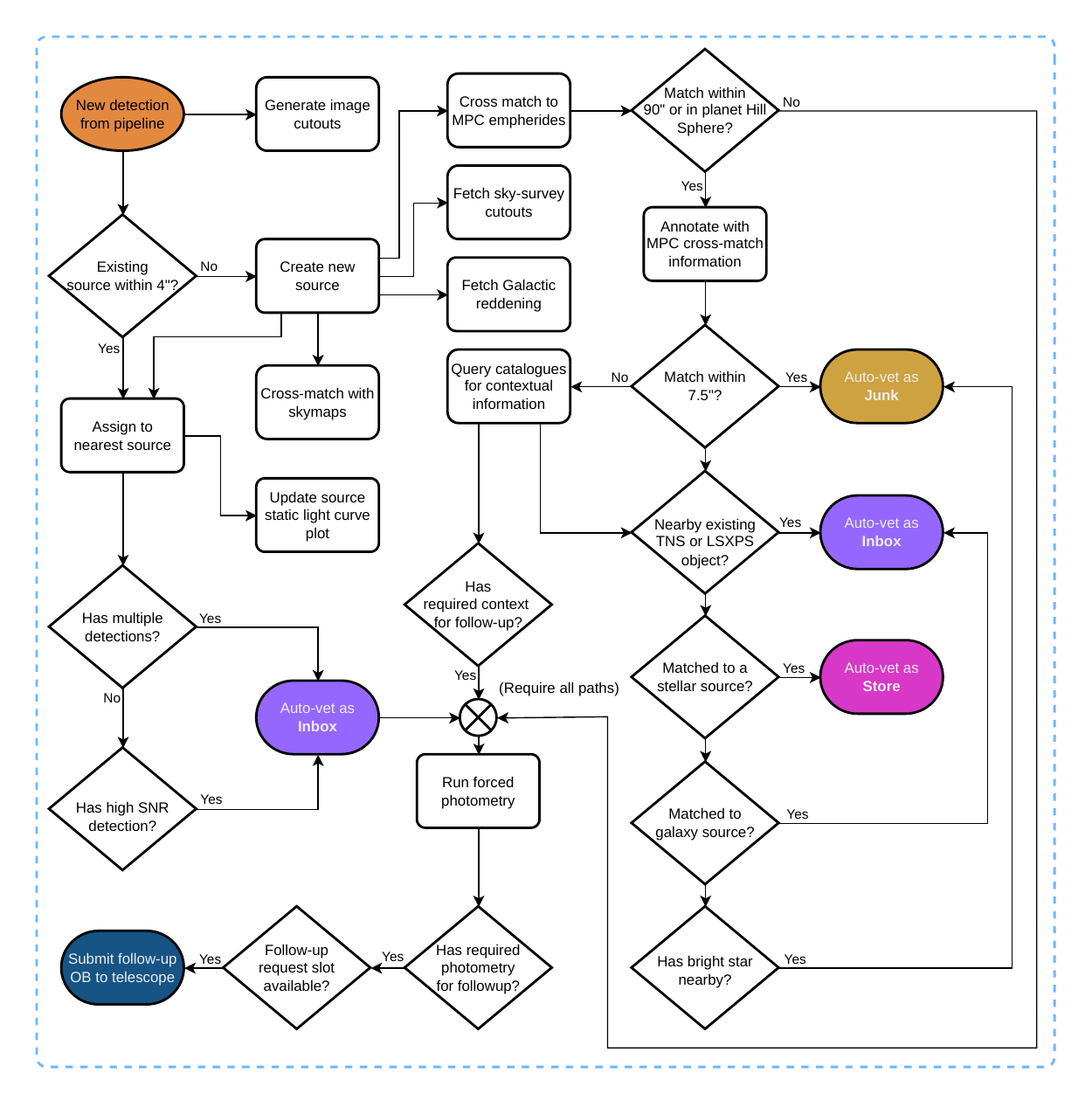}
    \caption{Schematic visualisation of the processing performed upon ingestion of a new difference image source into the GOTO marshall. Fuller details on these processes are in \cref{sec:marshall_ingest,sec:marshall_vetting,sec:marshall_forcedphotometry,sec:marshall_followup}. The summing junction ($\bigotimes$) must have all input paths followed by a source for the flow to continue (i.e. forced photometry is only performed if a source 1) was auto-vetted as \gloss{Inbox} due to high SNR or multiple detections, 2) has the required context for follow-up \textit{and} 3) is not matched with a minor planet or a planet's Hill Sphere. Colours are arbitrary and simply highlight the start and important end stages of the flow.)
    }
    \label{fig:goto_marshall_flowchart}
\end{figure*}

Ingestion of a new image causes a signal to add a task to the Celery queue to generate a JPEG preview of the whole image. Ingestion of new detections within the image follows a series of checks and tasks that are inserted into the Celery queue for processing in parallel or sequence, as required. A schematic overview of this process is shown in \cref{fig:goto_marshall_flowchart}, and described here.

\begin{enumerate}
    \item Check existing detections from the same image to ensure none with the same RA, and Declination coordinates is already held by the marshall database. Without this check, other methods of fetching photometry (see \cref{sec:marshall_forcedphotometry}) could conceivably ingest duplicates.
    \item Perform a `get or create' operation for a source, to either associate the detection with an existing source, or create a new one. For this, a row-level lock is acquired on a table of HEALPix \citep{healpix} indices covering those at and surrounding the detection coordinates, which prevents race conditions for source association or creation. The following are tried in order, ending early on success and then releasing the table lock:
    \begin{enumerate}
        \item  Return an exact match for the source at the exact RA, Declination coordinates.
        \item Return the nearest existing source in a $4^{\prime\prime}$ cone-search around the coordinates.
        \item Create a new source at the detection coordinates
    \end{enumerate}
    \item Generate JPEG cutouts around the detection from the \gloss{difference} image FITS file, producing cutouts for the \gloss{set}, \gloss{template} and \gloss{difference} data.
    \item Query the \kadmilos\ database to find the constituent \gloss{single} images used to create the \gloss{set} image and similarly generate cutouts of those.
    \item If a new source was created during ingestion of this detection:
    \begin{enumerate}
        \item Check if spatial-temporal coordinates are coincident with a known moving Solar System body (\cref{sec:marshall_moving_bodies}).
        \item Determine source context using {\sc Moriarty} \citep[][Killestein et al. in prep]{killstein_thesis}, which queries a range of catalogues to determine source disposition.
        \item Fetch stamps at the source position using the cutout services of optical sky-surveys SDSS DR12 \citep{sdss_dr12}, \panst\ DR2 \citep{panstarrs_dr2}, SkyMapper DR4 \citep{skymapper_dr4}, and Legacy Survey DR10 \citep{legacy_survey}.
        \item Perform automatic vetting procedures (\cref{sec:marshall_vetting}).
        \item Check if the source meets the criteria for automated follow-up (\cref{sec:marshall_goat}).
            \end{enumerate}
    \item Update or create the static light curve plot of the source with the new detection's photometry.
\end{enumerate}

After the above ingestion and processing steps, sources are then ready to be presented to collaboration members within the marshall, and are assessed and interacted with as deemed necessary, e.g. manual vetting and reporting (\cref{sec:marshall_human_vetting}) or triggering of follow-up (\cref{sec:marshall_followup}).

\subsection{Vetting processes}
\label{sec:marshall_vetting}

The marshall is chiefly a platform to vet candidate detections into respective categories or source lists (\cref{sec:marshall_sourcelists}). A stream of detections filtered on realbogus score from a model will still be impure, containing ambiguous detections or artefacts, as well as astrophysically real sources unrelated to transients. Human sifting therefore remains the final step before community reporting of transients, although several automated processes and additional filtering ease the task. 

\subsubsection{Source lists}
\label{sec:marshall_sourcelists}

Organisation of sources is primarily on source lists within the marshall: broad categorical tagging to focus human effort. The lists and short descriptions are given in \cref{tab:marshall_sourcelists}.

\begin{table}
\centering
\begin{tabular}{ p{1.2cm}  p{6.4cm} }
 List & Description \\ 
 \hline
 \gloss{Pending} & Ambiguous or unclear sources. All new sources created initially begin here.\\  
 \gloss{Inbox} & Sources deemed to require manual, human vetting. This is the primary list monitored by users for transient identification.\\
 \gloss{Stream} & Transient sources such as supernovae, afterglows and related objects.\\
 \gloss{Store} & Persistently varying or out-bursting sources, comprising mostly Galactic sources as well as Active Galactic Nuclei.\\
 \gloss{Junk} & Ephemeral detections such as image artefacts, satellite trails, and moving objects. New detections nearby these positions will not be matched to existing sources in this list, meaning future detections are treated as independent new sources.\\
 \gloss{Banish} & Detections in permanently compromised sky locations, such as the wings of bright stars. New detections at these positions will be matched to existing sources in this list to prevent the need for repeated vetting.\\
 \hline
 
\end{tabular}
\caption{Description of source lists used in the GOTO marshall vetting process.}
\label{tab:marshall_sourcelists}
\end{table}

All new ingestions enter the marshall as a \gloss{Pending} source list. The rate of entries into \gloss{Pending} is typically too numerous to be sifted manually. Sources are therefore automatically vetted based on several criteria.

\subsubsection{Automatic vetting}
\label{sec:marshall_autovetting}

The automated vetting decisions are shown as part of \cref{fig:goto_marshall_flowchart}, and they rely on either external contextual information for the source, or the nature of its detections by GOTO.

Any source matched to a known Solar System minor body is vetted as \gloss{Junk} (see \cref{sec:marshall_moving_bodies} for more details). Sources matched to a stellar source in either of the \textit{Gaia} \citep{gaia} or AAVSO International Variable Star Index \citep{aavso_vsx} catalogues are vetted as \gloss{Store}. A source with either a match in the Transient Name Server\footnote{\url{https://www.wis-tns.org/}} (TNS) or Living \textit{Swift} XRT Point Source Catalogue\footnote{\url{https://www.swift.ac.uk/LSXPS/}} \citep[LSXPS;][]{lsxps} are vetted to the \gloss{Inbox} for human inspection and reporting, as necessary. Should a source lie within 1\,arcmin of a galaxy catalogued by either the Galaxy List for the Advanced Detector Era+ \citep[GLADE+;][]{gladeplus} or NASA Extragalactic Database Local Volume Sample \citep[NED LVS;][]{nedlvs} it is also promoted to the inbox.

Beyond contextual classification, a source may be automatically vetted based on the results of its detection(s). If multiple difference image detections are associated to the source, or it has a single detection with $\textrm{SNR} > 15$, it is sent to the \gloss{Inbox}.

\subsubsection{Identification of moving-bodies}
\label{sec:marshall_moving_bodies}

Frequent visitors in time-domain survey difference images are moving objects. Since they are spatially displaced between the epochs of template and science observation, they manifest as astrophysically real sources of flux in difference images and can confound transient searches. Although they will appear at any given position only once, and so can be effectively mitigated by awaiting another epoch of observation for the same patch of sky, in the era of rapid transient identification and characterisation, it is desirable to identify objects in single epochs of detection.

High-proper motion stars are one such moving-body, although the number of such stars with angular velocities sufficient to displace more than a typical GOTO PSF width in the maximal few year epoch difference between science and template observations is small. Additionally, these objects are typically nearby, bright stars making such sources clear to identify by human eyeballers. Contextual information providers can also accommodate such objects trivially in labelling schemes given the exquisite proper motion constraints available from Gaia for sources down to the typical depth of GOTO. Given their low incidence rate, manual eyeballing of these sources is relied upon for vetting when they appear in the \gloss{Inbox} currently, although automated contextual labelling is being implemented.

Artificial satellites also pose as interlopers, with JWST a typical culprit. Such objects are typically bright, $L < 17$\,mag, and unassociated with any source on sky (stars or galaxies). As the sky density of such sources is low, and they are distinctive in their brightness and environment, they can effectively be managed by manual cross-checks of satellite positions for suspicious detections.

Solar-system bodies are numerous, and move on angular speeds ranging from marginal displacement between successive epochs in GOTO to significantly streaking within a single exposure. In the latter case, these are not a major contaminant for human vetters, since the exposure stacking minimises the presence of streaks in the combined \gloss{set} images, and the realbogus classifier {\sc gotorb} \citep{killestein2021} is further adept at ensuring they are not promoted as an object of interest for transient searches. In the parameter range of speeds where the bodies appear point-like within $\sim$minute-long exposures, but move significantly between successive epochs (hours-days), there is the most difficulty in distinguishing them from genuine transients. For these cases, we rely on the ephemeris service maintained by the Minor Planet Center\footnote{\url{https://www.minorplanetcenter.net/iau/mpc.html}} (MPC). The MPC provides orbital parameters and optical magnitude estimates for $\sim10^5$ Solar-system minor-bodies. Given the high rate of candidates requiring checking per night -- $O(10^{3.5})$ -- we perform the calculation of ephemerides locally. The package \pympc\ \citep{pympc} was developed to perform this task, and its implementation is detailed in \cref{app:pympc}. The output of \pympc\ is used to perform the following actions on new sources based on a minor or major Solar-system spatial match:

\begin{itemize}
    \item Matches of minor bodies brighter than GOTO's detection limit within $7.5^{\prime\prime}$ of the source's position are automatically vetted as \gloss{Junk} and not promoted to users.
    \item The nearest match within $90^{\prime\prime}$, if one exists, is displayed as contextual information for a source.
    \item A cautionary note is added if the source is within the Hill sphere of a major Solar-system body.
\end{itemize}

A source must have no nearby minor body, nor be within a major body's Hill Sphere, to proceed for checking of automatic follow-up criteria (\cref{sec:marshall_goat}).

\subsubsection{Manual vetting and reporting}
\label{sec:marshall_human_vetting}

\begin{figure}
    \centering
    \includegraphics[width=\linewidth]{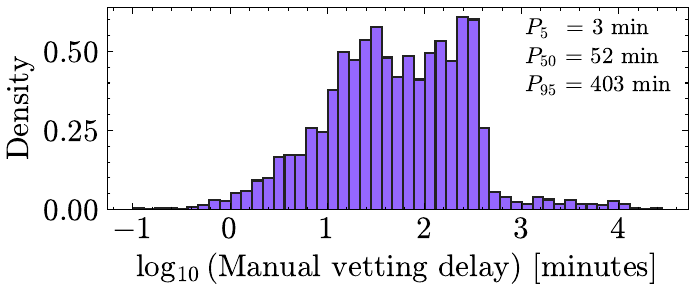}
    \caption{The time-delay between marshall auto-vetting placing a source into the \gloss{Inbox} and the \textit{final} manual vetting action performed on a source for sources in \cref{tab:vetting_numbers}. Note that sources may have been vetted multiple times in this window. Examples include being sent to the \gloss{Stream}, then moved to the \gloss{Store} once a spectroscopic classification reveals the source to be of that disposition, or to \gloss{Junk}/\gloss{Pending} if future observations of the location raises questions about its realness. Such cases contribute the small number with a delay of $>1$\,weekend.}
    \label{fig:marshall_vetting_delays}
\end{figure}

Collaboration members primarily check the \gloss{Inbox}, to make final vetting decisions on those identified by the auto-vetting decisions as worthy of consideration as real. One option is to return the source to the \gloss{Pending} source list, at which point it will be resurrected to the \gloss{Inbox} if new photometry meets the \gloss{Inbox} criteria again. The breakdown for destinations of \gloss{Inbox} sources is shown in \cref{tab:vetting_numbers}. Typically $\sim100$ objects require vetting per 24\,hour period, easily manageable by collaboration members. \cref{fig:marshall_vetting_delays} shows the delay-time distribution of manual vetting.

Once a source is identified as a candidate transient and sent to the \gloss{Stream} source list, the marshall initiates a forced photometry (\cref{sec:marshall_forcedphotometry}) task to gather the full history of variability at the location. The results of this forced photometry are then used to provide the discovery detection and recent non-detection when reporting the source to the TNS, which is performed via a form in the web interface, or externally via a Gamma-ray Coordination Network circular\footnote{\url{https://gcn.nasa.gov/}}.

\begin{table}
\centering
\begin{tabular}{llrr}
\hline
Origin & Destination & N per 24\,h & Percentage \\
\hline
\multicolumn{4}{c}{\textbf{Auto Vetting}} \\
\gloss{Pending} & \gloss{Inbox}   & 123   &  1.2\% \\
\gloss{Pending} & \gloss{Store}   & 1610  & 16.3\% \\
\gloss{Pending} & \gloss{Junk}    & 8158  & 82.5\% \\

\hline
\multicolumn{4}{c}{\textbf{Final Vetting}} \\
\gloss{Inbox} & \gloss{Stream}    & 17    & 14.2\% \\
\gloss{Inbox} & \gloss{Store}     & 23    & 18.4\% \\
\gloss{Inbox} & \gloss{Pending}   & 17    & 13.5\% \\
\gloss{Inbox} & \gloss{Banish}    & 2     &  1.5\% \\
\gloss{Inbox} & \gloss{Junk}      & 64    & 52.4\% \\
\hline
\end{tabular}
\caption{Breakdown of automatic vetting decisions by the marshall (\cref{sec:marshall_autovetting}), and the final destination of sources sent to the \gloss{Inbox} -- i.e. those which have been further checked by collaboration members (\cref{sec:marshall_human_vetting}). Numbers are rounded, and given for an average 24\,hour period from sources ingested over the time-frame 2025-06-01 to 2025-07-10.}
\label{tab:vetting_numbers}
\end{table}

\subsection{External trigger cross-matching}
\label{sec:marshall_crossmatching}

GOTO responds to different alerts at scheduler level -- currently GW, GRB and neutrino alerts from a variety of space missions and detectors. The same package is responsible for notifying the telescope scheduling software to react to a trigger is polled by the marshall to perform its own, independent cross-matching of sources and observations to triggered events.

\subsubsection{The need for an independent cross-matching system}
Sources discovered in images that were observed as part of a follow-up campaign of an alert have the alert's event name assigned to them by \kadmilos, which is pulled to the marshall, allowing users to filter for potential counterparts to the event. For a variety of reasons it is of interest that the marshall performs an independent cross-match of sources with potentially-linked skymap localisations of events, regardless of the telescope scheduler's observing reason when it discovered a source. These are:

\begin{itemize}
    \item The time between event triggers and the release of the alert may be delayed by minutes to hours. During that delay, sources discovered during serendipitous coverage of the localisation by GOTO would be missed.
    \item Scheduled follow-up is linked to at most one-event, whereas alerts may overlap spatio-temporally and so a given source should be identified as a potential counterpart to more than one alert.
    \item Particularly for GW events, both the localisation and electromagnetic science interest of events can change via alerts after updated analysis by the GW detector teams. Events may not warrant explicit follow-up at the time but are interesting to cross-match later, even with only serendipitous coverage.
\end{itemize}

\subsubsection{Alert ingestion and processing}
New alerts are polled regularly from the GOTO scheduling database which handles alert parsing \cite[{\sc G-TeCS};][]{dyer_thesis,dyer_scheduling}. If an alert relates to a new event, relevant trigger information is stored in the marshall database and a skymap is extracted from the alert and assigned to it. The cross-matching system builds upon the Gravitational wave Electromagnetic RecoveRY code \citep[{\sc GERry};][]{gerry} and {\sc healpix-alchemy} \citep{healpix_alchemy}, accessed via underlying \postgres\ rangeset data types. These rangesets are used to store both Multi-order Coverage (MOC) skymaps \citep{MOC2014,MOC2019,MOC2022} and image \healpix\ pixel information. This approach enables rapid querying of source positions and image coverage against a skymap. 

All skymaps, source positions and image footprints are ingested into a specific cross-matching DB. In the case of GW events, distance information is also stored alongside the \healpix\ values. Whilst typical skymaps released by LVK in GW alerts are in the MOC format, which primarily allow for a large reduction in the number pixels to describe the skymap, skymaps released by other facilities in the GRB domain use flat resolution maps with $\sim100$ times the number of pixels. To alleviate the storage and computational cost, any flat resolution skymap is converted to a MOC prior to ingest via a modified implementation of \texttt{bayestar\_adaptive\_grid} in the {\sc ligo.skymap} package \citep{bayestar}\footnote{\url{https://lscsoft.docs.ligo.org/ligo.skymap/index.html}}. The probability density function is first estimated by interpolating the flat skymap using pixel coordinates and pixel probability values. The repeated rounds of upsampling draw from this interpolation until the maximum resolution of the MOC matches the resolution of the original skymap. Typically the process results in a MOC skymap with $\sim$3000 pixels corresponding to a $>$98\% reduction in overheads, without sacrificing the accuracy of the localisation contours.

Not all alerts provide a skymap. In particular, instruments capable of arcminute localisations, such as \textit{Swift's} X-ray Telescope \citep{swiftxrt} or Einstein Probe's Wide-field X-ray Telescope \citep{einsteinprobe}, report only coordinates and a localisation error radius in alert files. For these events a flat high resolution skymap is created assuming a 2D Gaussian probability distribution at the coordinates, with a spread matching the localisation error given in the alert. In these cases, only pixels inside a 5-sigma contour of the Gaussian centre are ingested.

If another alert and/or skymap is released for the same event, the event values are updated and the new skymap is ingested, which is set as the currently active skymap for the purposes of cross-matching. Sources already cross-matched with a previous skymap version, but are not matched to the active skymap, are flagged with a caution.

Coordination of survey effort during GW alert follow-up by surveys like GOTO has been formalised in platforms such as the GW Treasure Map\footnote{\url{https://treasuremap.space/}} \citep{treasuremap}, with an aim to avoid duplicated effort, and maximise recovery chances \citep{treasuremap2}. Although GOTO participates via submission to the GW Treasure Map platform, the scheduler is not influenced or aware of other surveys' pointings via any mechanism, and reacts in an isolated fashion to each alert.

\subsubsection{Source and image cross-matching}
New sources in the marshall are ingested into the cross-matching database, ready to be cross-matched with external triggers.
Alerts are initially marked as `high' and `low' priority, assigning, respectively, an immediate or interval cross-matching schedule by the marshall. GRB, neutrino and high significance GW alerts with prospects for an EM counterpart are all given `high' priority.

If there are active events designated as `high' priority, new sources are cross-matched immediately as they are ingested. For `low' priority alerts, source cross-matching is performed by a dedicated task at 15 minute intervals, which also carries out image cross-matching and 2D localisation coverage calculations, following \citet{gerry}. If a source is cross-matched with an event skymap, the joint source-skymap properties such as contour level and absolute magnitude (in the cases where the skymap pixels have distance information) are calculated and stored in a junction table between the source and skymap tables. Users can then filter cross-matched sources based on these properties. Events have `active' status and are cross-matched for fixed windows of time post-event: 3 and 5 days for GRB and GW events, respectively. After this, events are labelled `archived' with no regular cross-match updating. Events that do not qualify for automatic cross-matching (i.e. are immediately labelled as `archived' by the marshall) can have cross-matching run manually using the marshall interface.

\subsubsection{Promising sources}
The cross-matched sources streams are regularly parsed to identify the most promising optical counterpart candidates for both GW and GRB events. Sources in the \gloss{Stream}, \gloss{Inbox} and \gloss{Pending} source lists that were discovered within 24\,hour of the trigger time are required to have multiple detections post-trigger, with no pre-trigger detections from forced photometry (\cref{sec:marshall_forcedphotometry}). 

Depending on the type of event, identification of a promising source follows different routes. For GW events, sources are marked as promising if they have been matched to a galaxy which has a distance estimate consistent with the GW distance estimate at the source position, and if the source has a peak absolute magnitude that is very generously consistent with a kilonova -- a wide range of $-10 \geq M_L \geq -17$\,mag is used, to allow for uncertain distances and photometry systematics. 

For GRB events, a source is marked as promising if the light curve can be fit with a power law decay with an index $-0.3\geq\alpha\geq-3$, which encompasses theoretical and observational constraints on the optical behaviour of GRBs \citep[e.g.][]{grb_optical_afterflows}. If a promising source is identified by either means, the source is highlighted visually in the marshall UI, automatically appears at the top of any cross-match source lists and triggers a team Slack alert for collaboration members.

\subsection{Forced Photometry}
\label{sec:marshall_forcedphotometry}

For the generation of sources, the marshall relies on the ingestion of photometry of automatically extracted sources (following \cref{sec:single_source_detection}) in \gloss{difference} images. This process is however incomplete for marginal detections, which may not pass the source detection criteria, nor is it able to provide non-detections for transient sources. For this, we rely on a forced photometry routine, which mimics the aperture-corrected photometry automatically extracted, and uses the same calibration products for the respective image. The positions of apertures are instead selected irrespective of the image properties, and are rather set by the source(s) one wishes to measure. 

The implementation of forced photometry has been extracted from the \kadmilos\ infrastructure and now lives as a stand-alone package accessible via an API or a web-frontend to handle both users' and automated requests (e.g. from the marshall), independent of the pipeline. Full details of the implementation are deferred to Jarvis et al. (in prep).

\subsection{Triggering of follow-up facilities}
\label{sec:marshall_followup}

The marshall has the capability to trigger other facilities via user-interaction, or automatically once a source meets pre-defined criteria (\cref{sec:marshall_goat}). Currently Liverpool Telescope \citep{liverpool_telescope} imaging, spectroscopy \citep{sprat} and polarimetry \citep{moptop}, and pt5m \citep{pt5m} imaging are implemented.

\begin{figure}
    \centering
    \includegraphics[width=\linewidth]{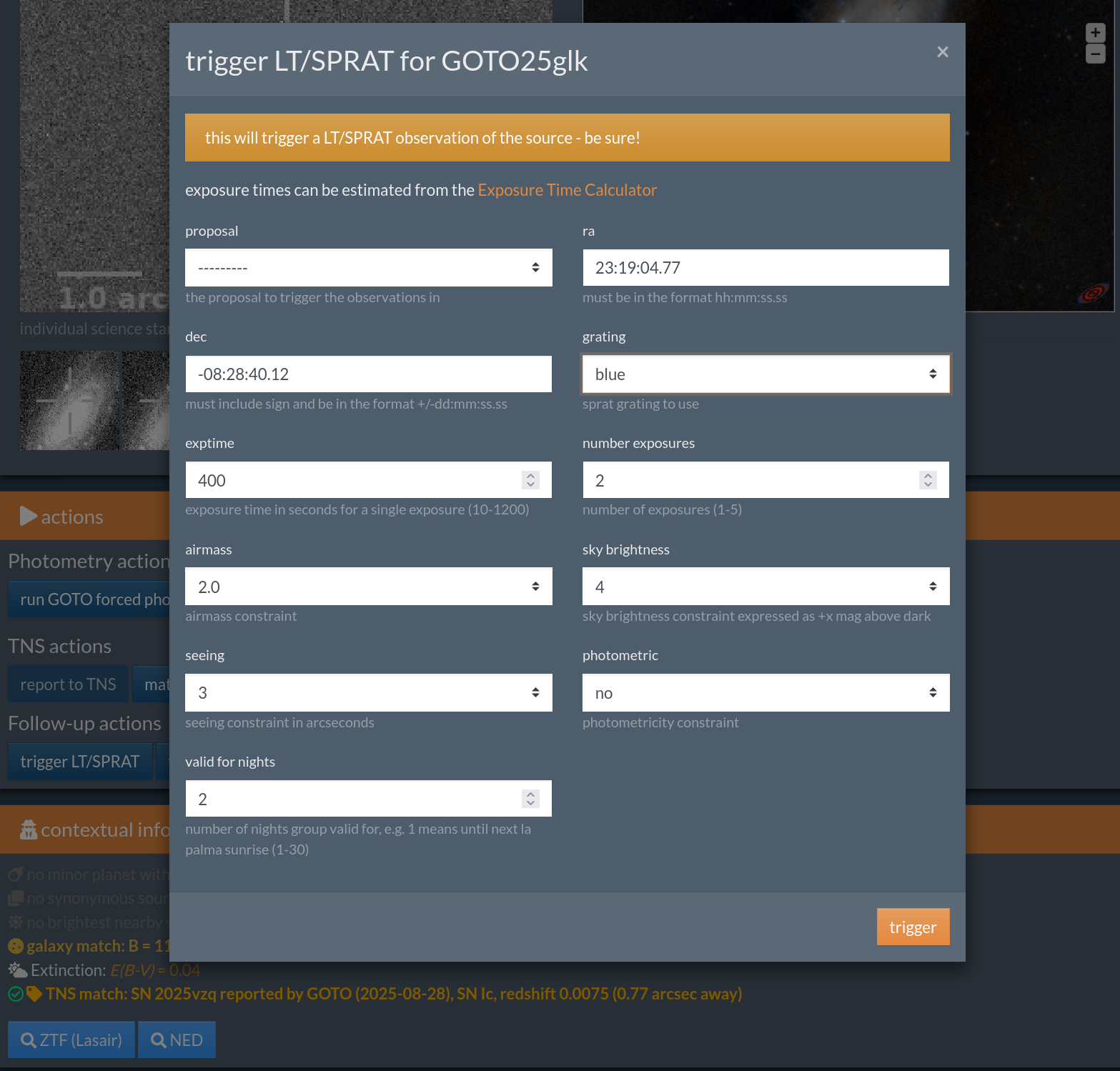}
    \caption{An example webform in the GOTO marshall allowing a user to trigger an LT spectroscopic observation using the SPRAT (Spectrograph for the Rapid Acquisition of Transients) instrument.}
    \label{fig:marshall_sprat}
\end{figure}

Users can define the observation sequence for each instrument via a web-form, which is then sent to the relevant facility (\cref{fig:marshall_sprat}). Additional integration of facilities with APIs is ongoing.

\subsubsection{Automatic follow-up scheduling}
\label{sec:marshall_goat}

The GOTO Auto-Trigger (GOAT) is a system within the marshall which performs a series of checks on new sources, based on both contextual and photometric information (\cref{fig:goto_marshall_flowchart}), to determine if the source is viable for automated follow-up by another facility. These checks performed by the GOAT on each new source arriving into the marshall are described in \cref{tab:trigger_criteria}. The variables shown are tuned per-follow-up facility, to reflect the external telescopes' capabilities.

\begin{table*}
\centering
\caption{Summary of triggering criteria checks applied to candidate transient sources. $\theta$ values refer to separations, $t$ values are times, and $m$ are magnitudes.}
\label{tab:trigger_criteria}
\begin{tabular}{p{3cm} p{6.5cm} p{6.5cm}}
\hline
\textbf{Criterion} & \textbf{Condition} & \textbf{Description} \\
\hline
\multicolumn{3}{l}{\textit{Source metadata and sky position}} \\
\hline
Vetting result & $\mathrm{vetting\_result} \in \{\mathtt{Stream}, \mathtt{Inbox}\}
$ & Exclude sources where auto-vetting has not indicated they are a viable transient candidate. \\
Galactic coordinates & $|b| > b_{\min} \;\; \mathbf{OR} \;\; |\ell - 180^\circ| < 90^\circ$ & Avoid crowded Galactic plane regions. \\
Declination & $\delta_{\min} < \delta < \delta_{\max}$ & Restrict to telescope-accessible declinations. \\
Reddening & $E(B - V) \leq E(B - V)_{\max}$ & Exclude regions with high Galactic extinction. \\
\hline
\multicolumn{3}{l}{\textit{Single catalogue cross-matches}} \\
\hline
Nearest Minor planet & $\theta_{\rm MP} \geq \theta_{\rm MP, lim} \;\; \mathbf{OR} \;\; m_{\rm MP} \geq m_{\rm MP,lim}$ & Avoid triggering on known minor planets. \\
Planet Hill sphere & No match & Reject sources coincident with planetary Hill spheres. \\
Bright nearby star &\(
\begin{array}{ll}
\textbf{NOT}& (m_{\rm bright} < 9 \;\; \mathbf{AND} \;\; \theta_{\rm bright} < 25^{\prime\prime}) \\
\textbf{NOR} \;\; & (m_{\rm bright} < 14 \;\; \mathbf{AND} \;\; \theta_{\rm bright} < 7.5^{\prime\prime})
\end{array}
\) & Avoid diffraction spikes or bright star residuals. \\
Gaia match & $\theta_{\it Gaia} > 2.5^{\prime\prime} \;\; \mathbf{UNLESS} \;\; \theta_{\rm galaxy} < 10^{\prime\prime} $ & Avoid triggering on {\it Gaia} stellar sources which are not located near to a catalogued galaxy (where galaxy nuclei or point-like components could exist). \\
\hline
\multicolumn{3}{l}{\textit{Moriarty contextual information}} \\
\hline
Moriarty classification & ${\rm moriarty\_classification} \notin \{\mathtt{var}, \mathtt{agn}\}$ & Avoid known stellar variable and AGN sources. \\
Candidate host association & $\theta_{\rm galaxy} < \theta_{\rm galaxy,lim}$ \;\; \textbf{OR} \;\; located in a galaxy cluster & Require some extragalactic context. \\
\hline
\multicolumn{3}{l}{\textit{Light-curve and photometric behaviour}} \\
\hline
Recent robust detection & $\exists \, i \in \mathcal{D}_{\rm det} : t_i > t_{\rm now} - \Delta t_{\rm lim} \;\; \textrm{and} \;\; \mathrm{RB}_i \ge \mathrm{RB}_{\rm lim}$ & Ensures there exists at least one recent, high realbogus entry in the set of detections. \\
Number of detections & 
\makecell{
$\mathrm{count} \{ $\\
$\;\;\;\;\;\;\;\;\;\;  i \in \mathcal{D}_{\rm det} : t_i > t_{\rm now} - \Delta t_{\rm lim} \;\; \textrm{and} \;\; \mathrm{SNR}_i \ge \mathrm{SNR}_{\min} $\\
$\} \ge N_{\rm det,\min}$
}
& Ensure a minimum number of recent, high SNR entries in the set of detections. \\
Historical detections & $\mathrm{count} \left\{ i \in \mathcal{D}_{\rm det} : t_i \leq t_{\rm now} - \Delta t_{\rm lim} \right\} \le N_{\rm hist,\max}$ & Exclude sources with historical entries in the set of detections (e.g. variables, AGN, persistent difference artefacts). \\
Non-detections & $\forall j \in \mathcal{D}_{\rm non} : t_j > t_{i_{\rm first}} \;\; \Rightarrow \;\; m_j < m_{i_{\rm first}}$ & Enforces all non-detection limits after the first detection are brighter than it, to ensure a rising light curve. \\
Latest detection age & $t_{i_{\rm latest}} > t_{\rm now} - 24$\,hr & Ensure source has been recently observed. \\
Brightness & $m_{i_{\rm latest}} < m_{\rm lim}$ & Latest magnitude must be bright enough to follow up. \\
Rise rate & $\dot{m} \leq \dot{m}_{\min}$ & Require a rise brightness rate, measured in magnitudes, and relative to non-detections or earlier detections. \\
\hline
\end{tabular}
\end{table*}

\begin{figure}
    \centering
    \includegraphics[width=\linewidth,clip=true,trim=0.9cm 0.3cm 0.9cm 0.4cm]{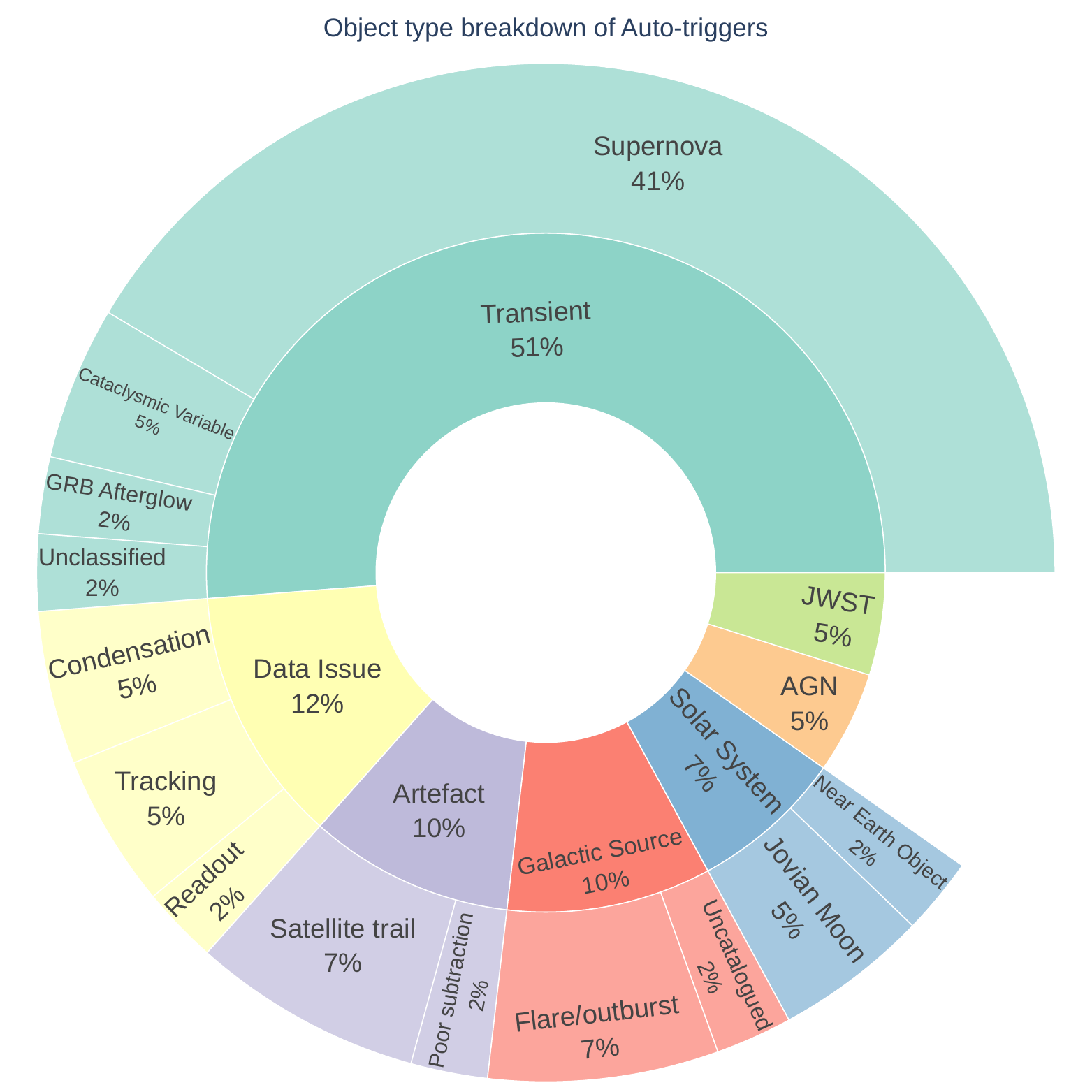}
    \caption{A breakdown of the nature of sources passing the GOAT checks (\cref{tab:trigger_criteria}) for automatic triggering of spectroscopic follow-up by LT/SPRAT. The triggers include those for sources passing early implementations of the checks, making it is a pessimistic view of the current performance (see \cref{sec:marshall_goat}).
    }
    \label{fig:goat_triggers}
\end{figure}

In \cref{fig:goat_triggers} we show a breakdown of the ultimate nature of sources that were triggered by the GOAT during 2025. Happily enough, even from a single detection, the checks select a majority of real astrophysical objects (transients, Galactic sources and AGN), although both image quality (both data issues during hardware acquisition and artefacts from software processing) and moving objects (Solar-system objects and JWST) contribute contaminants. 

It should be noted that this distribution includes triggers made during development of the final checking criteria, and so includes a number of triggers that would no longer be made in the current iteration. For example, early iterations did not include a check of whether a source was in a planet's Hill sphere, which allowed for Jovian Moon triggers when they are vexatiously in conjunction with catalogued galaxies. 

\begin{figure}
    \centering
    \includegraphics[width=\linewidth]{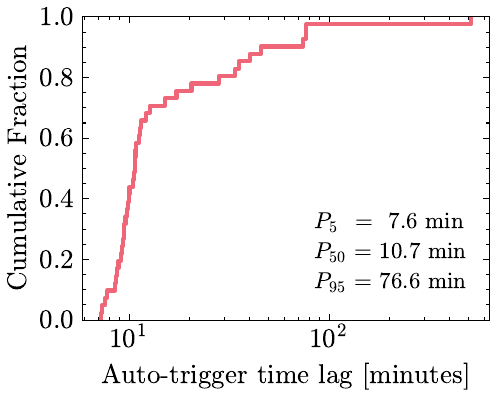}
    \caption{Cumulative delay-time distribution for LT/SPRAT triggers submitted by the GOAT. The time is measured from GOTO shutter close until the marshall sends the request to the LT API. Percentiles describing the distribution are shown in the plot as text.}
    \label{fig:goat_delays}
\end{figure}

\cref{fig:goat_delays} shows that the GOAT typically inserts observing sequences into the LT observation queue around 11\,minutes after shutter close. If they are picked up expediently by the LT scheduler, this offers a route to early characterisation of new discoveries \citep[e.g.][]{magee_iax}.


\subsection{Citizen Science integration}

\defcitealias{KilonovaSeekersI}{Killestein and Kelsey et al. \ \citeyear{KilonovaSeekersI}}

The marshall also has close integration with the Kilonova Seekers citizen science project \citepalias{KilonovaSeekersI}, which makes use of volunteer classifications of difference image stamps to identify various transients and outbursting sources in GOTO data \citep{GOTO0650}. The marshall regularly polls the Kilonova Seekers backend via a 
private API, awaiting new high confidence sources from the project. Once a source has at least eight votes cast by volunteers, with an agreement of $\ge80$\,percent that the source is astrophysically real, the marshall gains access to Kilonova Seekers information on the source.
This information includes names of a random subset of volunteers who voted the candidate as real, which allows subsequent TNS reporting of the source via the marshall to automatically populate a remark in the discovery report, providing credit for the volunteers' efforts.

Additionally, the Kilonova Seekers upload pipeline bootstraps contextual information from the marshall, enabling uploads to be rebalanced dynamically to maximise the number of probable transients sent to the platform -- specifically by not uploading transients already reported, and down-weighting minor planets in the random subset selected.

\subsection{Evaluation of transient recovery performance}
\label{sec:marshall_evaluation}

A high-level overview of the overall performance of the transient workflow in the GOTO project is made in this section, using reporting statistics from the public TNS database.

\begin{figure}
    \centering
    \includegraphics[width=\linewidth]{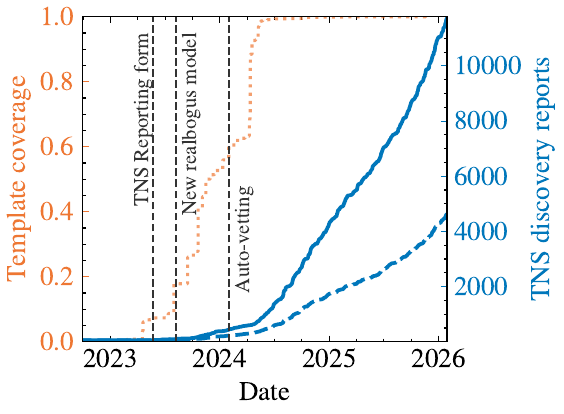}
    \caption{The cumulative sum of TNS discovery reports by GOTO (blue solid line), and the subset of first-discoveries (blue dashed line). The green dotted line shows the fraction of \gloss{template} images held by the pipeline -- a reasonable proxy of sky coverage fraction where difference imaging could be performed. Significant marshall milestones assisting the reporting of new transient candidates are also indicated.}
    \label{fig:goto_tns_stats}
\end{figure}

The TNS database shows 4623 first discoveries by GOTO between the dates of 2019-05-08 and 2026-01-28, with a total of $11800$ reported transient candidates in that same window.
The typical TNS reporting rate was $\sim2$\,reports/day until May 2024 when a $\sim~50$\,percent increase in template coverage (reaching almost complete coverage) and new marshall features (primarily Auto-vetting) were deployed. These combined to provide a step change in reporting to $15$--$16$\,reports/day, which has remained consistent until early 2026. The cumulative distribution of GOTO TNS reports is visualised in \cref{fig:goto_tns_stats}. The median magnitude of first detection magnitudes reported to the TNS for GOTO is $m_L\simeq19.5$\,mag over 2025, broadly consistent with expectations from the sky-survey depths (\cref{fig:set_image_depths}) when considering a more robust confirmation than a single $5\sigma$ detection is required to meet the threshold of reporting an object to the community.

The overall TNS reporting rate did not significantly alter in the periods of GW-detector down-time (namely Apr-Jun 2025 and Nov 2025-early 2026). Outside these windows, significant time was dedicated to covering GW localisations during the O4 observing runs of the GW detectors, in favour of the all-sky survey. There is the hint of an up-tick in the first-discovery rate since late 2025, when observing strategies to revisit regions of the sky at higher cadence have been implemented.

In addition to the overall transient discovery rate, the workflow has delivered confirmed optical afterglow detections for GRBs \citep{belkin2024,dimple2025,kumar2025}, including events initially localised only to degrees-scale regions. These recoveries, within hours of the high-energy triggers, exemplify the combination of wide-field observations via rapid robotic response, with low-latency data analysis and automated candidate vetting to efficiently isolate true optical counterparts from large candidate streams -- a practical validation of the marshall prioritisation strategy.

\section{Summary and outlook}

We have detailed the baseline infrastructure for transient discovery with GOTO in the preceding sections, and its performance and feature set has met the design requirements of the transients working group within the project:
robust detection and reporting of sources to the community starting just minutes after GOTO shutter close, with automatic follow-up being triggered for high priority sources. GOTO is discovering transients at a rate commensurate with prolific contemporary surveys such as ATLAS and ZTF, and adds valuable coverage to the global effort to map the changing night-sky on short time-scales.

Future upgrades for this transient workflow continue. Some key avenues for extracting additional robustness of the framework, and science exploitation of the discoveries have been identified:

\begin{itemize}
    \item Airflow 3.0 was released in April 2025, offering DAG versioning. An upgrade is planned which will help maintain provenance of GOTO outputs and a more tractable DAG history in the face of the slow-changing dimensions of GOTO hardware.
    \item The \textit{Gaia} Synthetic Photometry Catalogue \citep{gspc} offers a route to directly compute a calibration catalogue natively in the model GOTO-L filter bandpass, with a consequently smaller colour term required to be calculated per UT (to account for filter variations). An in-development online calibration process for forced photometry employs a similar technique using synthetic photometry and time-variable colour corrections to achieve increased accuracy of GOTO photometry (Jarvis et al. in prep).
    \item The bank of \gloss{template} images could be superseded by identification deeper stacks of multiple existing \gloss{set} images now that GOTO has visited its entire survey footprint many times. This would prevent the depth of the \gloss{template} images compromising the \gloss{difference} image depths in the face of improved throughput of the GOTO hardware and/or longer exposure-time sub-surveys.
    \item Software developed as part of the Vera C. Rubin LSST \citep{lsst} include advanced Solar-system and moving body packages \citep{sorcha}, which can be used for improved cross-checking of discoveries against known objects.
    \item Photometry and calibration of all sources in the science frames is not required for transient discovery. The differential in processing time arising from the density of sources (\cref{fig:single_and_set_timings}) could be removed by trimming the number of calibration and photometered sources per image down to a level where it does not materially affect the final results. In practice, however, the most crowded images are also likely those suffering from the most foreground extinction, and so are already of the least interest for transient searches. Although complete photometry is of interest for variable star searches, bespoke workflows for crowd-optimised photometry is required for robust results, and \kadmilos\ remains focused on the low-latency discovery of transients.
    \item Improvements to the contextual classifier, \textsc{Moriarty}, to include deep survey catalogues, redshift catalogues from large spectroscopic surveys \citep[e.g. ][]{desi_dr1}, and more robust galaxy host association, to minimise false associations.
    \item Deploying new machine learning models for real-bogus classification and more granular multi-task labelling, which run online to mitigate the effects of data/concept drift, and can be regularly fine-tuned with modest numbers of recent, high-quality labels.
    \item Source vetting currently is a final step in the marshall, whereas interesting behaviour can be observed in historic sources (late rebrightenings of transients, extreme behaviour of AGN etc.) Filters designed to place human eyes back on these sources can be added to the framework of the marshall.
    \item The current reporting method is via TNS discovery reports. In future, a model of releasing streams of alerts to which brokers can listen, akin to that employed by ZTF \citep{ztfalertstream} and LSST,\footnote{\url{doi:10.71929/rubin/2586493}} will further open the use of GOTO data to the wider community.
\end{itemize}

\section*{Acknowledgements}

The authors' thanks go to the contributors of the open source software projects upon which this work stands. Gratitude is given to David Young for fruitful conversations and inspiration on marshall development.

We additionally thank the two reviewers for constructive feedback and suggestions to improve the clarity of the work.

The Gravitational-wave Optical Transient Observer (GOTO) project acknowledges the support of the Monash-Warwick Alliance; University of Warwick; Monash University; University of Sheffield; University of Leicester; Armagh Observatory \& Planetarium; the National Astronomical Research Institute of Thailand (NARIT); Instituto de Astrof\'sica de Canarias (IAC); University of Portsmouth; University of Turku; University of Birmingham; and the UK Science and Technology Facilities Council (STFC, grant numbers ST/T007184/1, ST/T003103/1 and ST/Z000165/1).

JDL, DON and MP acknowledge support from a UK Research and Innovation Future Leaders Fellowship (grant references MR/T020784/1 and UKRI1062).
TLK acknowledges support from a Warwick Astrophysics prize post-doctoral fellowship made possible thanks to a generous philanthropic donation.
DS acknowledges support from The Science and Technology Facilities Council (STFC) via grants ST/T007184/1, ST/T003103/1, ST/T000406/1, ST/X001121/1.
BW acknowledges the UKRI’s STFC studentship grant funding, project reference ST/X508871/1.
JC acknowledges support by the Spanish Ministry of Science via the Plan de Generaci\'on de Conocimiento through grant PID2022-143331NB-100.
DLC acknowledges support from the Science and Technology Facilities Council (STFC) grant number ST/X001121/1.
LK acknowledges support for an Early Career Fellowship from the Leverhulme Trust through grant ECF-2024-054 and the Isaac Newton Trust through grant 24.08(w).
DMS acknowledges support through the Ram\'on y Cajal grant RYC2023-044941, funded by MCIU/AEI/10.13039/501100011033 and FSE+.
SMa acknowledges financial support from the Research Council of Finland project 350458.
SMo is funded by Leverhulme Trust grant RPG-2023-240.
AS acknowledges the Warwick Astrophysics PhD prize scholarship made possible thanks to a generous philanthropic donation.
RLCS acknowledges funding from Leverhulme Trust grant RPG-2023-240.

This research has made use of data and/or services provided by the International Astronomical Union's Minor Planet Center.

This work has made use of data from the European Space Agency (ESA) mission {\it Gaia} (\url{https://www.cosmos.esa.int/gaia}), processed by the {\it Gaia} Data Processing and Analysis Consortium (DPAC, \url{https://www.cosmos.esa.int/web/gaia/dpac/consortium}). Funding for the DPAC has been provided by national institutions, in particular the institutions participating in the {\it Gaia} Multilateral Agreement.

This publication uses data generated via the Zooniverse.org platform, development of which is funded by generous support, including a Global Impact Award from Google, and by a grant from the Alfred P. Sloan Foundation.

Computing facilities were provided by the Scientific Computing Research Technology Platform of the University of Warwick.

This research has made use of the NASA/IPAC Extragalactic Database, which is funded by the National Aeronautics and Space Administration and operated by the California Institute of Technology, including NED LVS (10.26132/NED8).

Funding for SDSS-III has been provided by the Alfred P. Sloan Foundation, the Participating Institutions, the National Science Foundation, and the U.S. Department of Energy Office of Science. The SDSS-III web site is \url{http://www.sdss3.org/}.
SDSS-III is managed by the Astrophysical Research Consortium for the Participating Institutions of the SDSS-III Collaboration including the University of Arizona, the Brazilian Participation Group, Brookhaven National Laboratory, Carnegie Mellon University, University of Florida, the French Participation Group, the German Participation Group, Harvard University, the Instituto de Astrofisica de Canarias, the Michigan State/Notre Dame/JINA Participation Group, Johns Hopkins University, Lawrence Berkeley National Laboratory, Max Planck Institute for Astrophysics, Max Planck Institute for Extraterrestrial Physics, New Mexico State University, New York University, Ohio State University, Pennsylvania State University, University of Portsmouth, Princeton University, the Spanish Participation Group, University of Tokyo, University of Utah, Vanderbilt University, University of Virginia, University of Washington, and Yale University. 

The national facility capability for SkyMapper has been funded through ARC LIEF grant LE130100104 from the Australian Research Council, awarded to the University of Sydney, the Australian National University, Swinburne University of Technology, the University of Queensland, the University of Western Australia, the University of Melbourne, Curtin University of Technology, Monash University and the Australian Astronomical Observatory. SkyMapper is owned and operated by The Australian National University's Research School of Astronomy and Astrophysics. The survey data were processed and provided by the SkyMapper Team at ANU. The SkyMapper node of the All-Sky Virtual Observatory (ASVO) is hosted at the National Computational Infrastructure (NCI). Development and support of the SkyMapper node of the ASVO has been funded in part by Astronomy Australia Limited (AAL) and the Australian Government through the Commonwealth's Education Investment Fund (EIF) and National Collaborative Research Infrastructure Strategy (NCRIS), particularly the National eResearch Collaboration Tools and Resources (NeCTAR) and the Australian National Data Service Projects (ANDS).

The Pan-STARRS1 Surveys (PS1) and the PS1 public science archive have been made possible through contributions by the Institute for Astronomy, the University of Hawaii, the Pan-STARRS Project Office, the Max-Planck Society and its participating institutes, the Max Planck Institute for Astronomy, Heidelberg and the Max Planck Institute for Extraterrestrial Physics, Garching, The Johns Hopkins University, Durham University, the University of Edinburgh, the Queen's University Belfast, the Harvard-Smithsonian Center for Astrophysics, the Las Cumbres Observatory Global Telescope Network Incorporated, the National Central University of Taiwan, the Space Telescope Science Institute, the National Aeronautics and Space Administration under Grant No. NNX08AR22G issued through the Planetary Science Division of the NASA Science Mission Directorate, the National Science Foundation Grant No. AST–1238877, the University of Maryland, Eotvos Lorand University (ELTE), the Los Alamos National Laboratory, and the Gordon and Betty Moore Foundation.

The Legacy Surveys consist of three individual and complementary projects: the Dark Energy Camera Legacy Survey (DECaLS; Proposal ID \#2014B-0404; PIs: David Schlegel and Arjun Dey), the Beijing-Arizona Sky Survey (BASS; NOAO Prop. ID \#2015A-0801; PIs: Zhou Xu and Xiaohui Fan), and the Mayall z-band Legacy Survey (MzLS; Prop. ID \#2016A-0453; PI: Arjun Dey). DECaLS, BASS and MzLS together include data obtained, respectively, at the Blanco telescope, Cerro Tololo Inter-American Observatory, NSF's NOIRLab; the Bok telescope, Steward Observatory, University of Arizona; and the Mayall telescope, Kitt Peak National Observatory, NOIRLab. Pipeline processing and analyses of the data were supported by NOIRLab and the Lawrence Berkeley National Laboratory (LBNL). The Legacy Surveys project is honored to be permitted to conduct astronomical research on Iolkam D'uag (Kitt Peak), a mountain with particular significance to the Tohono O'odham Nation.

NOIRLab is operated by the Association of Universities for Research in Astronomy (AURA) under a cooperative agreement with the National Science Foundation. LBNL is managed by the Regents of the University of California under contract to the U.S. Department of Energy.

This project used data obtained with the Dark Energy Camera (DECam), which was constructed by the Dark Energy Survey (DES) collaboration. Funding for the DES Projects has been provided by the U.S. Department of Energy, the U.S. National Science Foundation, the Ministry of Science and Education of Spain, the Science and Technology Facilities Council of the United Kingdom, the Higher Education Funding Council for England, the National Center for Supercomputing Applications at the University of Illinois at Urbana-Champaign, the Kavli Institute of Cosmological Physics at the University of Chicago, Center for Cosmology and Astro-Particle Physics at the Ohio State University, the Mitchell Institute for Fundamental Physics and Astronomy at Texas A\&M University, Financiadora de Estudos e Projetos, Fundacao Carlos Chagas Filho de Amparo, Financiadora de Estudos e Projetos, Fundacao Carlos Chagas Filho de Amparo a Pesquisa do Estado do Rio de Janeiro, Conselho Nacional de Desenvolvimento Cientifico e Tecnologico and the Ministerio da Ciencia, Tecnologia e Inovacao, the Deutsche Forschungsgemeinschaft and the Collaborating Institutions in the Dark Energy Survey. The Collaborating Institutions are Argonne National Laboratory, the University of California at Santa Cruz, the University of Cambridge, Centro de Investigaciones Energeticas, Medioambientales y Tecnologicas-Madrid, the University of Chicago, University College London, the DES-Brazil Consortium, the University of Edinburgh, the Eidgenossische Technische Hochschule (ETH) Zurich, Fermi National Accelerator Laboratory, the University of Illinois at Urbana-Champaign, the Institut de Ciencies de l'Espai (IEEC/CSIC), the Institut de Fisica d’Altes Energies, Lawrence Berkeley National Laboratory, the Ludwig Maximilians Universitat Munchen and the associated Excellence Cluster Universe, the University of Michigan, NSF’s NOIRLab, the University of Nottingham, the Ohio State University, the University of Pennsylvania, the University of Portsmouth, SLAC National Accelerator Laboratory, Stanford University, the University of Sussex, and Texas A\&M University.

BASS is a key project of the Telescope Access Program (TAP), which has been funded by the National Astronomical Observatories of China, the Chinese Academy of Sciences (the Strategic Priority Research Program ``The Emergence of Cosmological Structures'' Grant \#XDB09000000), and the Special Fund for Astronomy from the Ministry of Finance. The BASS is also supported by the External Cooperation Program of Chinese Academy of Sciences (Grant \#114A11KYSB20160057), and Chinese National Natural Science Foundation (Grant \#12120101003, \#11433005).

The Legacy Survey team makes use of data products from the Near-Earth Object Wide-field Infrared Survey Explorer (NEOWISE), which is a project of the Jet Propulsion Laboratory/California Institute of Technology. NEOWISE is funded by the National Aeronautics and Space Administration.

The Legacy Surveys imaging of the DESI footprint is supported by the Director, Office of Science, Office of High Energy Physics of the U.S. Department of Energy under Contract No. DE-AC02-05CH1123, by the National Energy Research Scientific Computing Center, a DOE Office of Science User Facility under the same contract; and by the U.S. National Science Foundation, Division of Astronomical Sciences under Contract No. AST-0950945 to NOAO.

Software: Apache Airflow\footnote{\url{https://github.com/apache/airflow}}, \texttt{astropy} \citep{astropyI,astropyII,astropyIII}, \texttt{lmfit} \citep{lmfit}, \texttt{astroscrappy} \citep{astroscrappy}, \texttt{sep} \citep{sep}, \texttt{astrometrynet} \citep{astrometrynet}, \texttt{q3c} \citep{q3c_paper}, \texttt{rbf}\footnote{\url{https://github.com/treverhines/RBF}}, \texttt{dill}\footnote{\url{https://github.com/uqfoundation/dill}}.

\section*{Data Availability}

 Underlying image-level data are to be released within the scope of GOTO data releases. Prior to this, reasonable requests to the corresponding author can be made. The high-level data required to reproduce figures is available upon request. A static copy of the main \kadmilos\ data pipeline code as described here is held at \url{https://github.com/GOTO-OBS/kadmilos-public}.

\section*{Conflict of Interest}

Authors declare no conflict of interest.



\bibliographystyle{rasti_shortauthorlist}
\bibliography{references} 




\appendix

\section{Additional figures}
\label{app:additional_figures}

\begin{figure*}
    \centering
    \includegraphics[width=\linewidth]{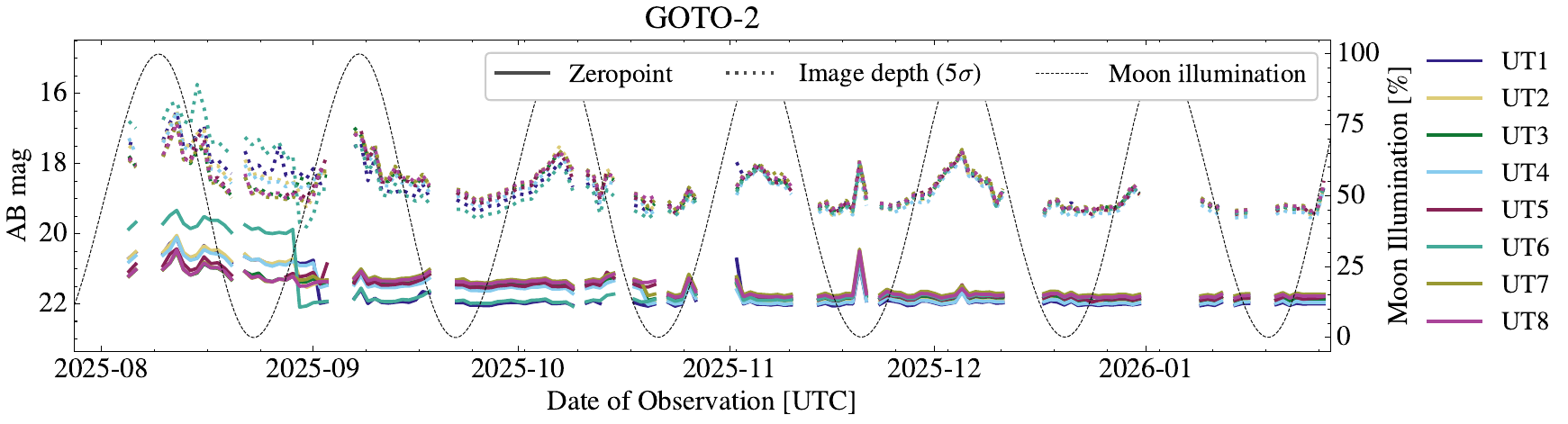}
    
    \vspace{1cm}
    
    \includegraphics[width=\linewidth]{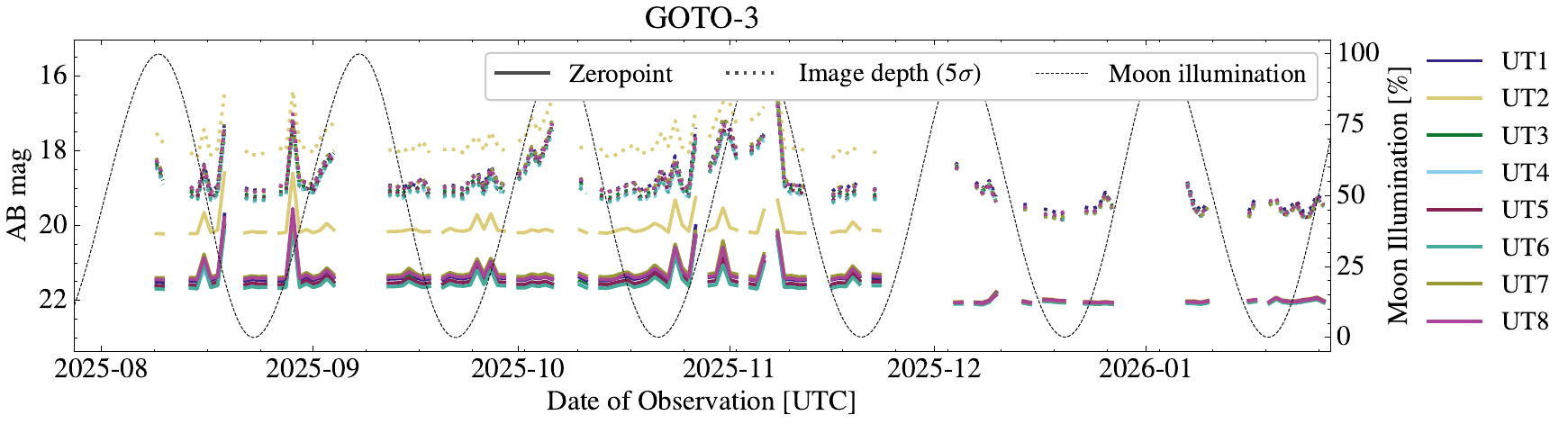}
    
    \vspace{1cm}
    
    \includegraphics[width=\linewidth]{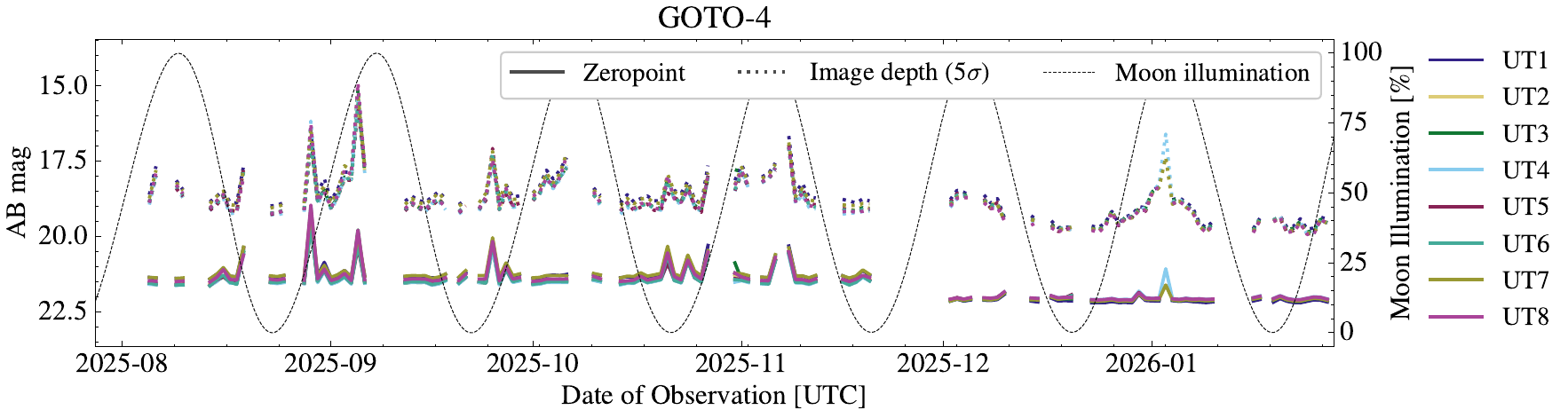}
    \caption{Same as \cref{fig:zeropoints_1} but shown here for mounts GOTO-2, GOTO-3 and GOTO-4.}
    \label{fig:zeropoints_other}
\end{figure*}

A visualisation of zeropoints and sensitivity of GOTO-[2-4] is shown analogously to \cref{fig:zeropoints_1} for GOTO-1 in \cref{fig:zeropoints_other}.

\begin{figure*}
    \centering
    \includegraphics[width=\linewidth]{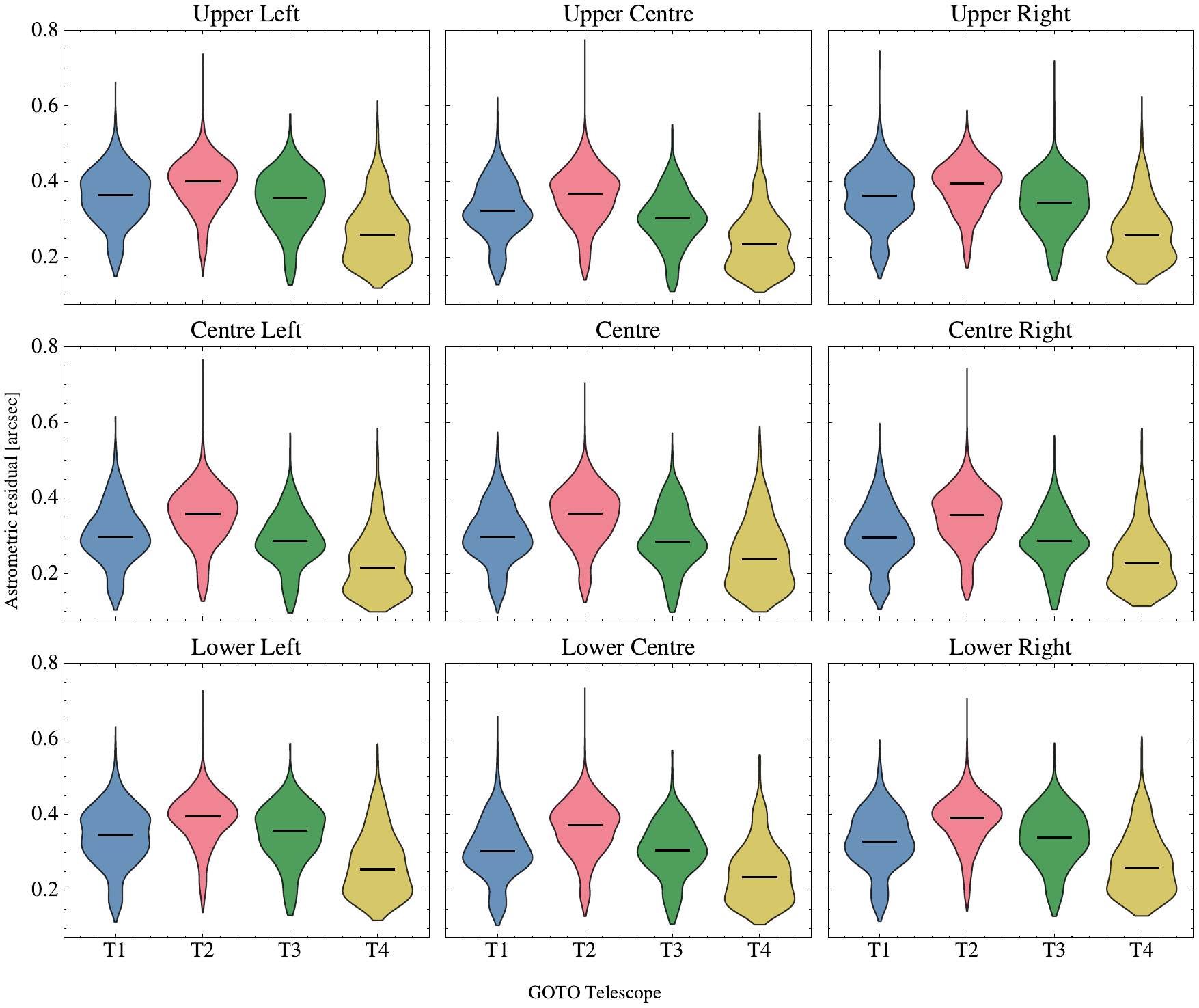}
    \caption{Astrometric residuals for each GOTO telescope in different regions of the image sensor. The regions are identical to the coloured heatmap regions in \cref{fig:astrometic_residual_heatmap}. Distributions are shown as kernel-density estimates, with median values as black horizontal bars.}
    \label{fig:astrometric_residual_violins}
\end{figure*}

Telescope-specific astrometric residual distributions, as additional insight into the distributions that produce the GOTO-level values shown in \cref{fig:astrometic_residual_heatmap} are shown in \cref{fig:astrometric_residual_violins}.

\section{\spalipy}
\label{app:spalipy}

\spalipy\ \citep{spalipy} provides detection-based alignment of astronomical images, allowing for a flexible transformation of a `source' image onto the pixel-coordinate system of a `template' image. It was built initially from an implementation of image alignment by Malte Tewes. \spalipy\ can perform its own detection routine to find common sources, or use externally supplied catalogues (e.g. output from SExtractor, \citealt{sextractor}). The catalogues are searched for matching ``quads'' following the algorithm of \citet{astrometrynet}. The best-matching quad in each image is used to derive an initial affine transformation that maps the `source' detections into the coordinates of the `template'. 

After applying the affine transform to all source detection coordinates, residual offsets to the template positions are evaluated. Because wide-field instruments, including GOTO, frequently exhibit spatially varying distortions, an affine transform is typically insufficient to achieve sub-pixel alignment across the field of view. To rectify this, \spalipy\ can optionally fits two independent 2D spline surfaces (\texttt{x} and \texttt{y} alignment residuals) that capture non-uniform, spatially dependent deformations. These spline corrections are then applied to generate a final, refined mapping of the source image onto the template’s frame.

\section{\pympc}
\label{app:pympc}

\pympc\footnote{\url{https://github.com/lyalpha/pympc}} \citep{pympc} is a package to calculate bulk positions of major and minor Solar-system bodies for requested epochs and spatial coordinates. It interfaces with data products of the Minor Planet Center (MPC) and uses functionality of the \pyehpem\ package\footnote{\url{https://rhodesmill.org/pyephem/}} to achieve this. \pympc\ is used in the GOTO data pipeline to determine if a known minor or major body is in the vicinity of a new detection in a difference frame, using the astrometric position of that detection and time of observation.

The data from MPC comprises the \texttt{mpcorb}, \texttt{nea}, and \texttt{comets} catalogues, which, provide orbital elements of minor bodies (asteroids, near-earth asteroids, and comets, respectively). These are parsed into the file format of {\sc XEphem}\footnote{\url{https://xephem.github.io/XEphem/Site/}}, with internal handling then done by \pyehpem\ to perform propagation of orbits. The underlying C-library ({\sc libastro}) used to propagate orbits provides only geocentric astrometric coordinates, and so \pympc\ then performs its own topocentric correction based on the position of the observer provided, typically via a MPC Observatory code. The propagation of the orbits is from the epoch defined in the MPC data to the requested epoch by the user. MPC epochs are updated typically every $\sim200$\,days, and so for checking of the current epoch, there can be a significant propagation required. The offline propagation provided by \pyehpem\ is not a full solution as performed by the MPC, and, owing to lack of treatment for planetary perturbations and other effects, the accuracy of a minor body's position is sensitively linked to the length of propagation required. Typically this is $\lesssim1^{\prime\prime}$ for epoch differences of several months or less, however it can sharply increase to $\sim$\,arcminute levels when far from an MPC epoch update.

Additionally, positions of the planets and their major moons, for which pre-defined objects exist in \pyephem, are also checked during a position search. Since this list of moons is incomplete, \pympc\ allows one to determine if a detection resides (in projection) within the Hill sphere of a planet at that epoch, in which case one should be cautious about the transient nature of the detection.

Typically searches (with multiprocessing) takes of order a few seconds - dominated almost entirely by the minor body search. In practice, one can readily cache their positions for a given epoch to make light work of searching many detections in a given set of images.


\bsp	
\label{lastpage}
\end{document}